\documentclass[12pt]{article}
\usepackage{amssymb,amsmath}
\usepackage{graphicx}
\usepackage{geometry}

\usepackage{natbib}

\usepackage[lined,boxed,ruled]{algorithm2e}

\usepackage[tight]{subfigure}
\newcommand{\I}[1]{\mathit{#1}}

\begin{document}

\title{Modeling self-organizing traffic lights with elementary cellular automata}
\author{Carlos Gershenson$^{1,2,3}$ and David A.\ Rosenblueth$^{1,2}$\\
$^{1}$ Instituto de Investigaciones en Matem\'aticas Aplicadas y en Sistemas \\
Universidad Nacional Aut\'onoma de M\'exico\\
Ciudad Universitaria\\
Apdo.\ Postal 20-726\\
01000 M\'exico D.F. M\'exico\\
Tel. +52 55 56 22 36 19 \
Fax +52 55 56 22 36 20 \\
\href{mailto:cgg@unam.mx}{cgg@unam.mx} \
\url{http://turing.iimas.unam.mx/~cgg} \\
\href{mailto:drosenbl@servidor.unam.mx}{drosenbl@servidor.unam.mx} \ \url{http://leibniz.iimas.unam.mx/~drosenbl/}\\
$^{2}$ Centro de Ciencias de la Complejidad \\
Universidad Nacional Aut\'onoma de M\'exico\\
$^{3}$Centrum Leo Apostel, Vrije Universiteit Brussel\\
Krijgskundestraat 33 B-1160 Brussel, Belgium
}
\maketitle

\begin{abstract}
There have been several highway traffic models proposed based on cellular automata. The simplest one is elementary cellular automaton rule 184. We extend this model to city traffic with cellular automata coupled at intersections using only rules 184, 252, and 136. The simplicity of the model offers a clear understanding of the main properties of city traffic and its phase transitions.

We use the proposed model to compare two methods for coordinating traffic lights: a \emph{green-wave} method that tries to optimize phases according to expected flows and a \emph{self-organizing} method that adapts to the current traffic conditions. The \emph{self-organizing} method delivers considerable improvements over the \emph{green-wave} method. For low densities, the \emph{self-organizing} method promotes the formation and coordination of platoons that flow freely in four directions, i.e.\ with a maximum velocity and no stops. For medium densities, the method allows a constant usage of the intersections, exploiting their maximum flux capacity. For high densities, the method prevents gridlocks and promotes the formation and coordination of ``free-spaces" that flow in the opposite direction of traffic.

\end{abstract}


\section{Introduction}

A mathematical model is an abstraction of a system.
Ideally, such an abstraction should be as simple as possible, provided
that the essential properties of the system are preserved. 
Hence, the complexity of a model faces a compromise between
simplicity and usefulness.
A complex model provides an accurate description of the system's
behavior, but may also bring added difficulty to
the representation, computation, and analysis of the model.
In the study of vehicular traffic, for example, numerous models have 
appeared in the literature \citep
{PrigogineHerman1971,Traffic95,Traffic97,Traffic99,Helbing1997,HelbingHuberman1998,ChowdhuryEtAl2000,Maerivoet:2005}.
Depending on the complexity of such models, they are useful for
different purposes.
Here we propose a vehicular model that is as simple as possible while
reproducing city traffic behavior. The purpose of our model is not predictive but explanatory.

Our model is based on elementary cellular automata, which we briefly present in the next section. This is followed by a summary of previous traffic models based on cellular automata. Our model is presented in Section \ref{sec:model}. We show results of a single intersection in Section \ref{sec:single}. We then focus on the problem of coordinating multiple traffic lights. We detail the mechanisms of a \emph{green-wave} method in Section \ref{sec:gw} and of a \emph{self-organizing} method in Section \ref{sec:so}. This is followed by results and a discussion of experiments carried out in a Manhattan-style simulation in Section \ref{sec:sims}. Potential refinements to the model and conclusions close the paper.

\section{Elementary cellular automata}

Cellular automata were perhaps first studied by Stanis{\l}aw
Ulam and John von Neumann \citep{vonNeumann1966} as a tool for modeling biological systems.
Later, John Conway devised his well-known ``Game of Life'' \citep{BerlekampEtAl82} using
such a formalism. 
More recently, the work of Stephen Wolfram \citeyearpar{Wolfram1986,Wolfram:2002} showed applications to many
areas of science, further increasing the interest in cellular automata.
As a modeling tool, cellular automata are especially valuable 
because of the simplicity of their specification, on the one hand, and
the complexity in their behavior, on the other hand.

An elementary cellular automaton is a collection of cells arranged on
a one-dimensional array.
A cell in such an
automaton has only two possible states (0 or 1, say).
Time is discrete and all cells' states are updated synchronously.
Moreover, the state of a cell in the next
time step, or ``tick'', depends only on the present states of that cell and those of
its nearest neighbors.
As a result, the behavior of an elementary cellular automaton can
be described by a table specifying the state a given cell
will have in the next ``generation'' based on 
the state of the cell to its
left, the state of the cell itself, and the state of the cell to its
right.
Such a table has as input these three current states and as output
the state of a cell in the next generation.
Wolfram names each elementary cellular automaton with the binary
numeral, called ``rule'',
resulting from reading the output of the table when the inputs are
lexicographically ordered.
Because cells' states are updated synchronously, an elementary cellular
automaton can be readily simulated with only two arrays of bits.

%

\section{Cellular automata models of vehicular traffic}
In this section, we first give an account of vehicular-traffic models
in general and then proceed to summarize models based on cellular
automata.

In empirical observations of highway traffic, it is possible to notice two
different regimes~\citep{Hall:1986}.
For low densities (number of vehicles per length unit), 
the flux (number of vehicles per time unit) shows an approximately linear behavior.
For higher densities, however, the flux exhibits strong
fluctuations resulting in a complex behavior, that
is still not clearly understood~\citep{ChowdhuryEtAl2000}.
First, such fluctuations prevent the use of a functional model.
Second, hysteresis has been noticed,
where the flux is greater when the density increases than when the
density decreases.
Third, metastable states (i.e.\ states in a precarious equilibrium)
have been observed. 
Authors normally distinguish between at least two different congested (``jammed")
regimes: the synchronized and the stop-and-go phases~\citep{ChowdhuryEtAl2000}.

As a result of this complexity,
myriad highway traffic models have appeared in the literature.
``Macroscopic'' models view traffic 
as a one-dimensional
compressible fluid~\citep{ChowdhuryEtAl2000}.
The ``microscopic'' approaches, by contrast, model each individual
vehicle.
Kinetic theories model traffic as a gas in which each particle
represents a vehicle~\citep{huang:87}.
The class of
``follow-the-leader'' models represents each 
vehicle with a
motion equation in a system of interacting classical
particles~\citep{ChowdhuryEtAl2000}. 
``Coupled-map lattice'' models treat time as a discrete variable and
the dynamical equations for each vehicle becomes a discrete dynamical
map~\citep{kaneko:93}. 
Finally, cellular-automata models play a prominent role.
One of the main reasons is that these models are computationally cheap.


Possibly the first traffic model that could be viewed as a cellular
automaton was that of Cremer and Ludwig~\citeyearpar{cremer:ludwig:86}.
By using binary one-dimensional arrays, this model
represents the presence/absence of a vehicle with
each of the two states of a cell, so that each vehicle occupies exactly
one cell.
These authors use bitwise Boolean
operations together with shifts to update the state of the automaton.
Successive applications of different such operations can simulate
acceleration, deceleration, lane changing, passing, and turning.
Observe that updating the state of each cell requires the state of
neighbors which are not the nearest ones to such a cell.
This cellular automaton, therefore, is not elementary.

Elementary cellular automata (ECA), nonetheless, can model traffic as well.
We can assume that whether a vehicle moves forward or not
depends only on the presence or absence of another vehicle
just in front.
Supposing that a vehicle moves one cell to the right if and only if
such a cell to the right is empty, then rule 184 corresponds to traffic
moving to the right \citep{Maerivoet:2005}.
Apparent movement is created as follows:
If the state of a cell is 1 and that of its right neighbor is 0, then
this rule assigns 0 to the cell's state in the next generation,
eliminating the vehicle from the current cell
(which accounts for 110 and 010).
Similarly, if the current cell's state is 0 and its left neighbor is
1, then the cell's next state is 1, thus apparently making the vehicle
in the left neighbor move one cell to the right
(covering cases 100 and 101).
Two other situations must be dealt with.
First, a vehicle cannot move if there is another vehicle right in front
(i.e.\ 111 and 011).
Second, if there is no vehicle in the cell or its left neighbor, then
in the next generation the cell will have no vehicle either
(i.e.\ 001 and 000).
See Table~\ref{table:ECArules}.

\begin{figure}
     \centering
     \subfigure{
          \label{fig:results_streetA}
          \includegraphics[width=.45\textwidth]{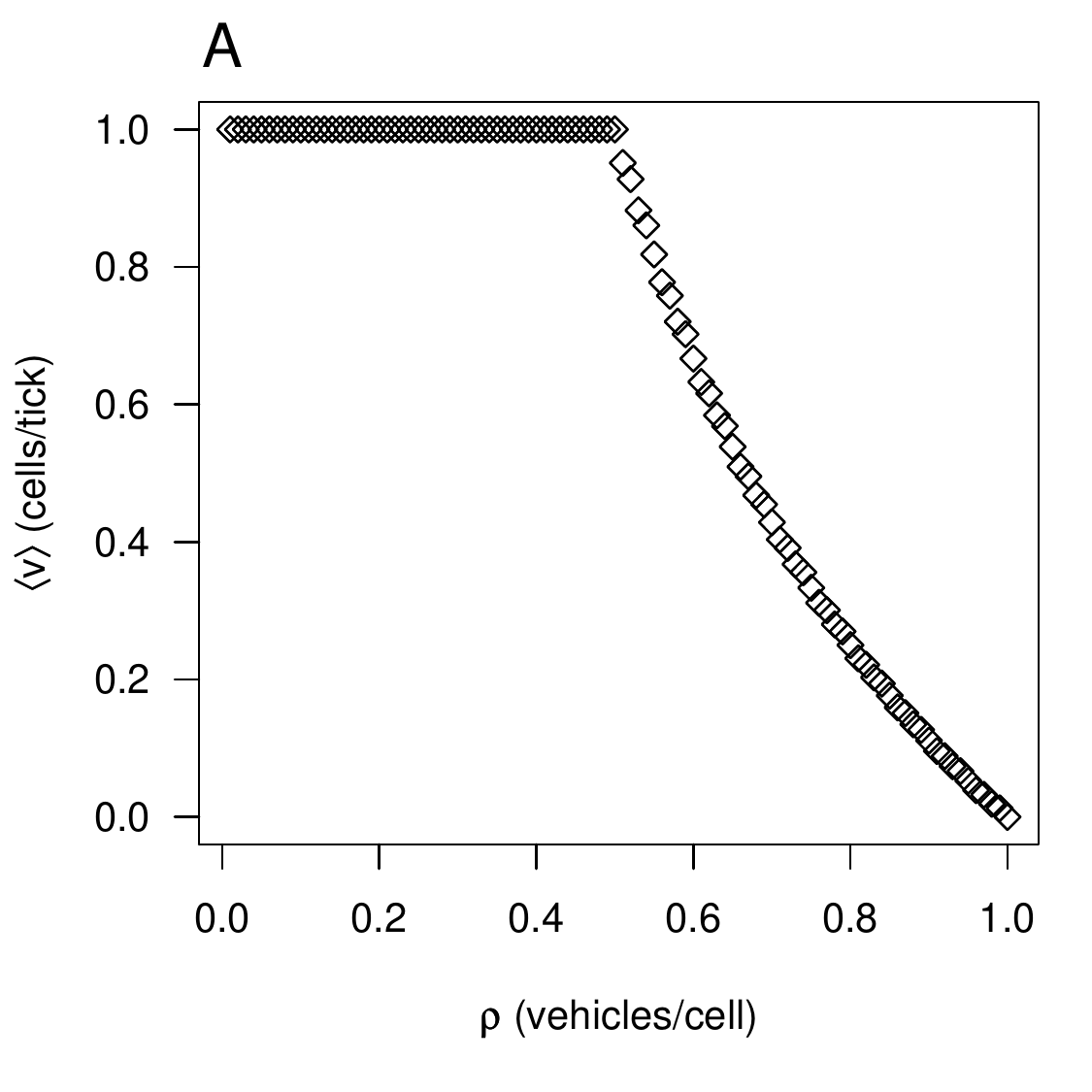}}
     \subfigure{
          \label{fig:results_streetB}
          \includegraphics[width=.45\textwidth]{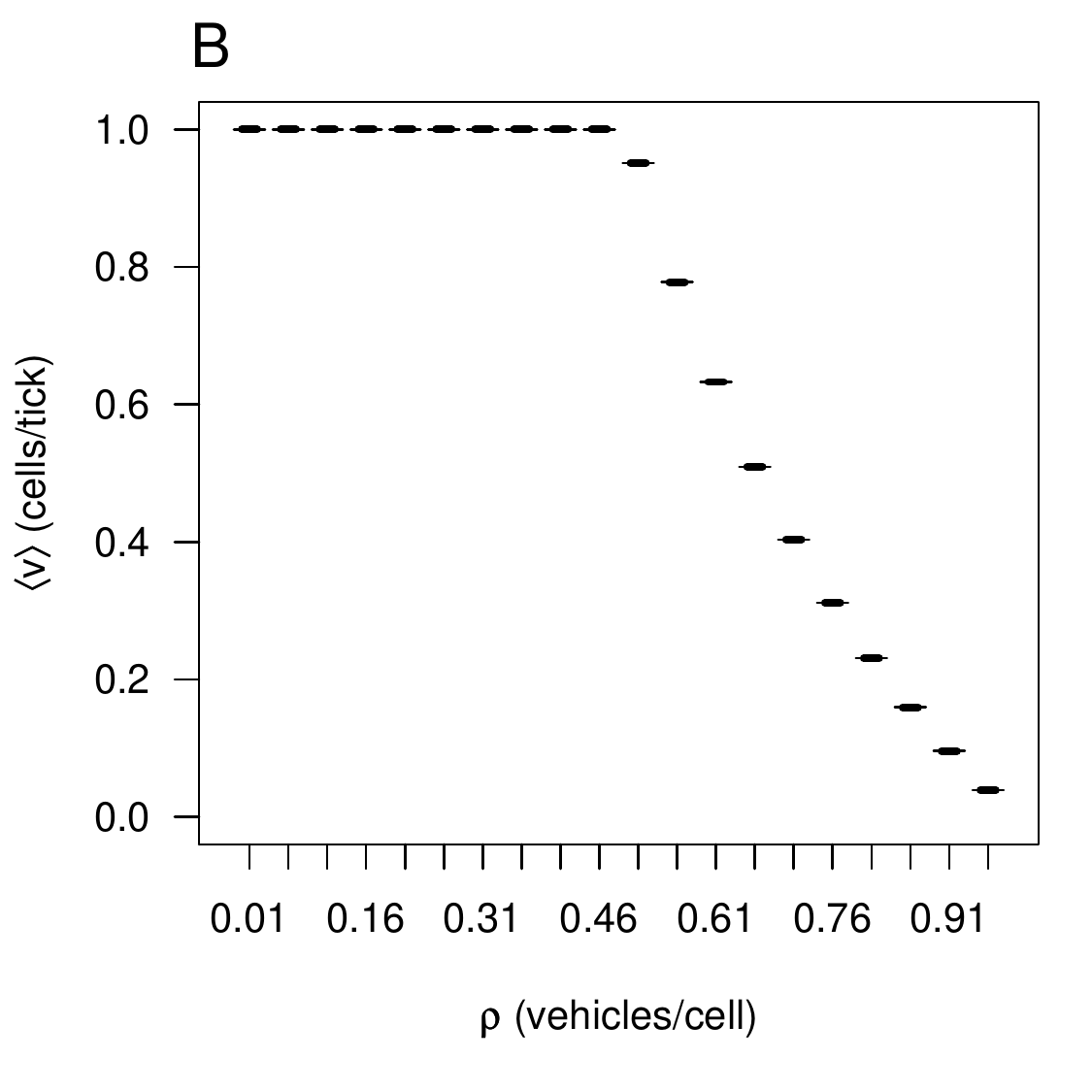}}
\\
     \subfigure{
          \label{fig:results_streetC}
          \includegraphics[width=.45\textwidth]{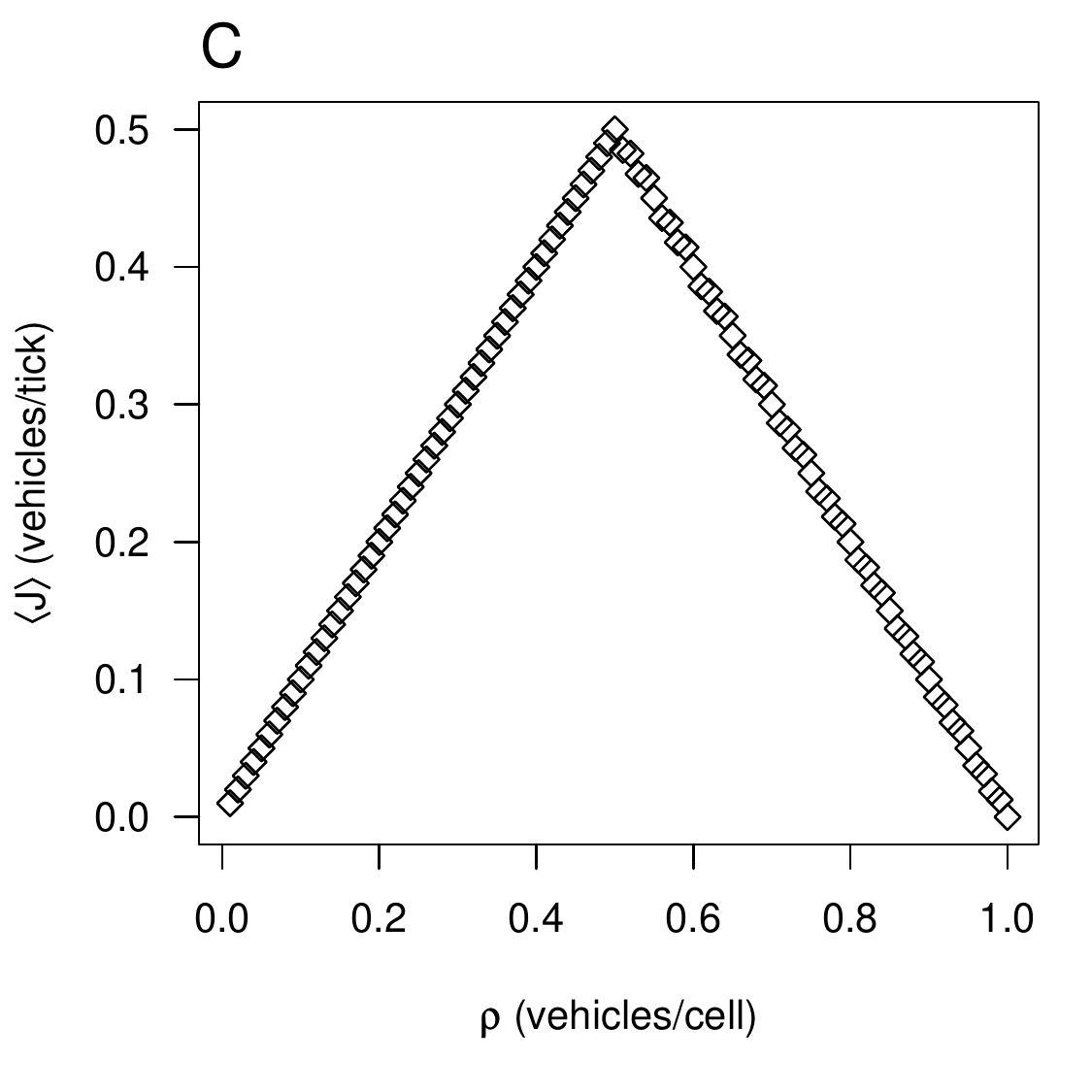}}
     \subfigure{
          \label{fig:results_streetD}
          \includegraphics[width=.45\textwidth]{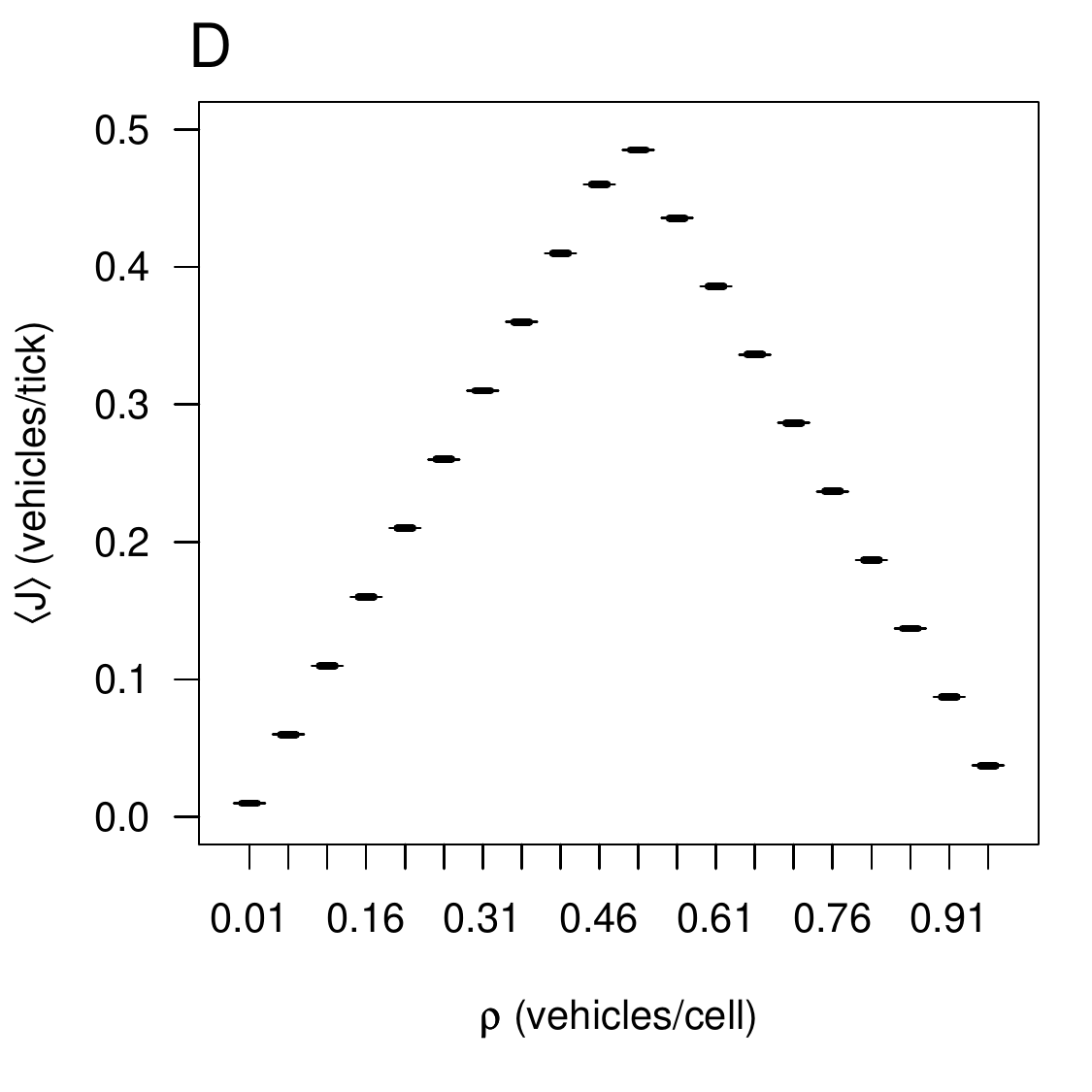}}

     \caption{Simulation results for rule 184: (A,B) average velocity $\langle v\rangle$ and (C,D) average flux $\langle J\rangle$ for different densities $\rho$: (A,C) single runs and (B,D) box plots of 50 runs per density.}
     \label{fig:results_street}
\end{figure}

\begin{figure}
     \centering
     \subfigure[]{
          \label{fig:184A}
          \includegraphics[width=.45\textwidth]{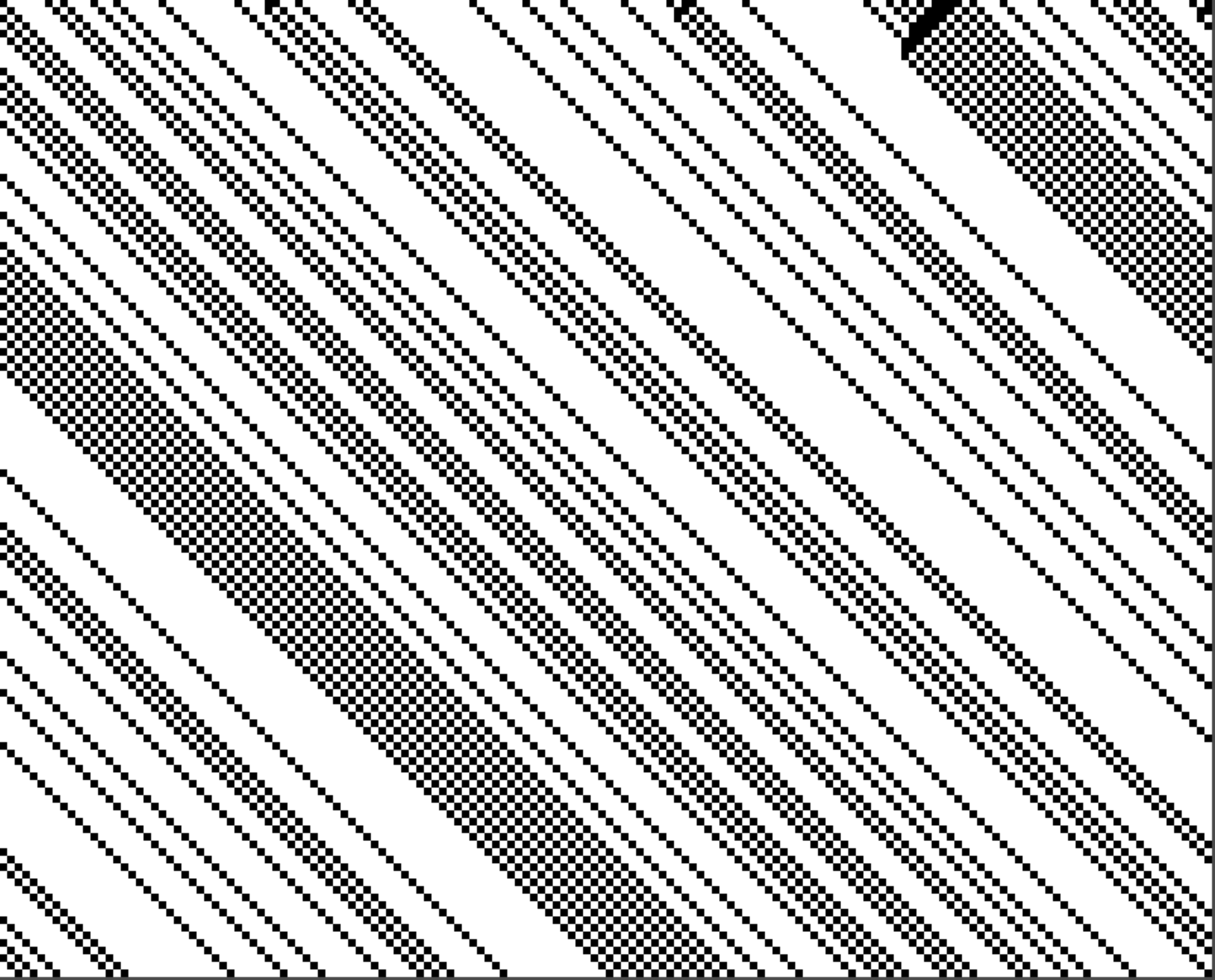}
	}
     \subfigure[]{
          \label{fig:184B}
          \includegraphics[width=.45\textwidth]{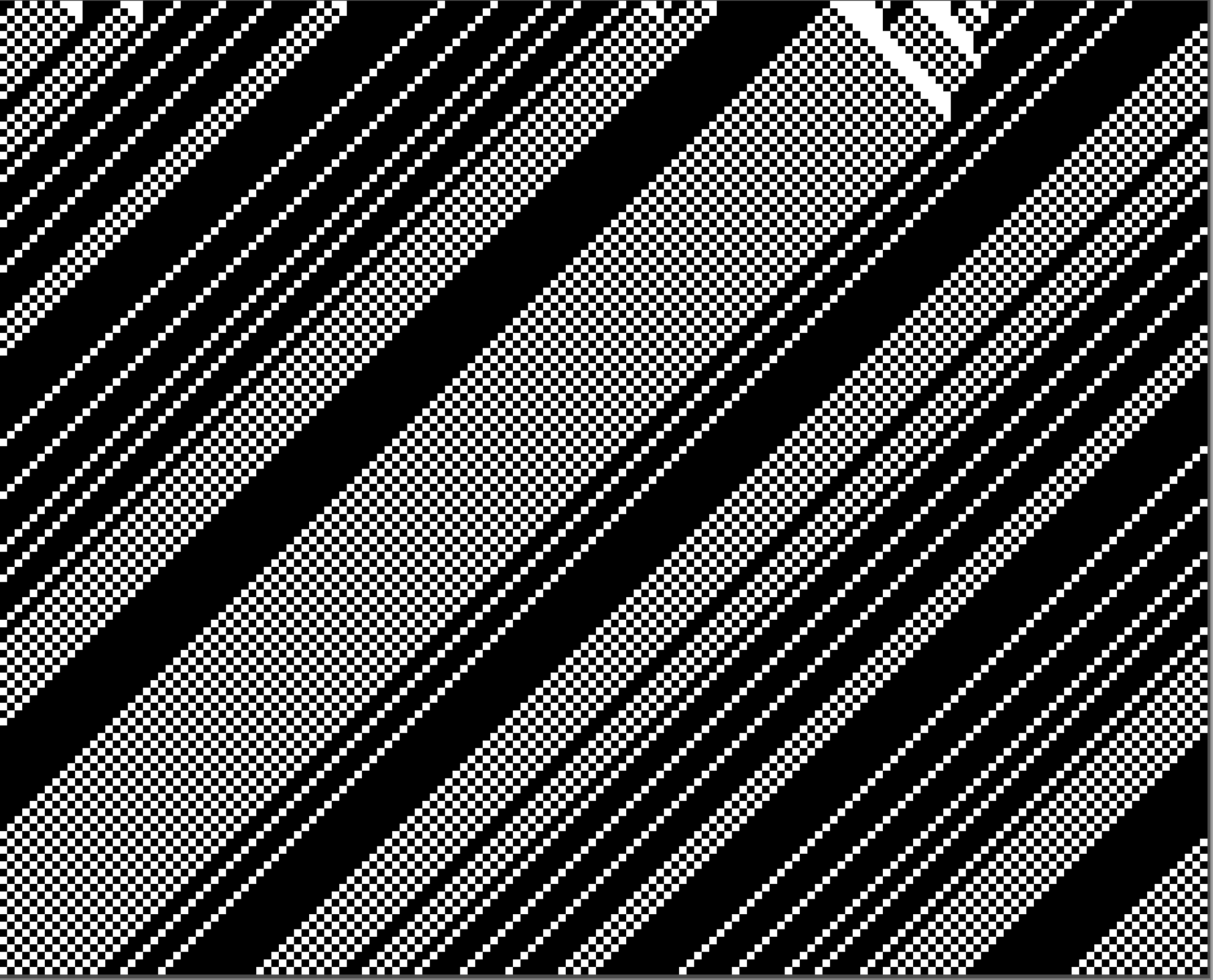}
	}

     \caption{Evolution of rule 184. Black cells (1) represent vehicles, white cells (0) represent spaces. Traffic flows to the right, time flows to the bottom: (A) In the \emph{free-flow} phase ($\rho=0.25$ shown) all vehicles flow at a velocity of one cell per tick. (B) In the \emph{jammed} phase ($\rho=0.75$ shown) jams move to the left, as vehicles can only advance when there is a free space ahead of them.}
     \label{fig:184}
\end{figure}

By randomly placing vehicles along,
this CA exhibits two kinds of behavior, i.e.\ phases,
depending on the initial density $\rho$ of vehicles.
For $\rho < 50\%$, vehicles flow freely, without
interacting with each other.
By contrast, for $\rho > 50\%$, the traffic flow is ``jammed'', as congestion waves move to the left, i.e.\ in the opposite direction of traffic. Average velocities and fluxes for different densities are shown in Figure \ref{fig:results_street}. The temporal evolution of rule 184 in both phases is shown in Figure \ref{fig:184}.

The traffic model of Nagel and Schreckenberg~\citeyearpar{NaSch1992} (NaSch) can
be seen as an elaboration of rule 184 with the following extensions:
(1) a variable (discrete) velocity associated with each vehicle, 
(2) acceleration (tending to attain the maximum velocity), 
(3) deceleration (due to the presence of other vehicles), and 
(4) a random tendency to slow down.
This slowdown attempts to model a human tendency to overreact when decelerating.
The NaSch model reproduces the appearance of spontaneous, also called
``phantom'', 
traffic jams.
Depending both on the global density and the probability to slow down, such
spontaneous traffic jams either disappear or survive indefinitely.

Note that a vehicle can have a velocity greater that one cell per tick.
This fact implies
that the NaSch traffic model, just as that of Cremer and Ludwig,
is a non-elementary cellular automaton. 
The reason is that the next state of a cell depends not only of those
of the immediate neighbors of such a cell, but also on those of other cells.

Many variants of the NaSch model have appeared, each with 
different degrees of realism. 
For example, Nagel and Paczuski~\citeyearpar{nagel:paczuski:95} 
inhibit the random slowdown for vehicles traveling at the maximum velocity.
This variation eliminates spontaneous traffic jams for the free-flow
regime by representing automatic cruise control.

Fukui and Ishibashi~\citeyearpar{fukui:ishibashi:96}, 
by contrast, only have a random behavior component for vehicles traveling
at the maximum velocity.
Such a random slowdown of maximum-velocity vehicles attempts to model the
fact that drivers traveling at high velocity (without cruise control)
cannot concentrate indefinitely.
In addition, accelerations in this model are instantaneous.


A two-dimensional cellular automaton modeling traffic was
devised by Biham, Middleton, and
Levine~\citeyearpar{biham:middleton:levine:92} (BML).
This model is interesting because of being remarkedly simple and yet
exhibiting self-organization as well as two distinct phases.
Cells form a bidimensional array.
Each cell can be viewed as representing either an empty space or a
vehicle which is traveling either to the right or upwards.
Except for the initial random placement of the vehicles (i.e., the
initial condition) this automaton is deterministic.
Boundary conditions are periodic, so that the number of vehicles of
each kind is preserved.
On even ticks, only upward-facing vehicles move, whereas on odd
ticks, only vehicles facing right do so, 
unless there is a nonempty space just in front.

Self-organization emerges when consecutive rows or columns have
vehicles moving one cell ahead or behind the next row or column, 
thus forming a diagonal of vehicles.
This pattern minimizes collisions and therefore maximizes speed.
Above a certain density, a global cluster appears which rapidly 
includes all vehicles, showing a ``sharp'' phase transition.

Biham, Middleton, and Levine~\citeyearpar{biham:middleton:levine:92} further
experimented with two variants of this model.
First, they obtained similar results when all vehicles were able to move
simultaneously; in case two tried to occupy the same cell one was
randomly chosen.
This showed that the observed behavior was not caused by the fact that
the vehicles in one direction took turns with those in the other
direction. 
Second, when two vehicles oriented in different directions were allowed
to occupy the same cell, the phase transition was gradual,
demonstrating that the sharp
phase transition was caused by the fact that a cell could not be
simultaneously occupied by two vehicles.

The BML model is interesting because of 
exhibiting complex behavior, 
but is not a realistic model of city
traffic.
More realistic city-traffic models have been developed as a common
generalization of the NaSch and BML
models~\citep{simon:nagel:97,chowdhury:schadschneider:99,schadschneider&:99,BrockfeldEtAl2001}.
Such generalizations
are essentially an extension of the BML model so
that streets have an arbitrary length (instead of only one cell), and
vehicles traveling in between crossings behave according to the NaSch
model.
Traffic lights are incorporated by making vehicles
decelerate or halt not only because of a vehicle
being in front, but also because of approaching a red traffic light.

In~\citep{schadschneider&:99}, the authors experimented with a regular
grid where vehicles move only to the right or upwards.
Traffic lights were synchronized and simultaneously alternate between
green and red (a traffic-light behavior which we call ``marching''~\citep{Gershenson2005}).
As in the BML model, there is a sharp phase transition between the \emph{free-flow}
and the \emph{jammed} regimes.

Another study combining both these models is that
of Brockfeld at al.~\citeyearpar{BrockfeldEtAl2001}.
This work compares the ``\emph{marching}" and ``\emph{green-wave}" (See Section \ref{sec:gw}) methods with random traffic lights.
The random traffic lights outperformed the other two behaviors.

%
%

Simon and Nagel \citeyearpar{simon:nagel:97} have developed a more elaborate combination of the NaSch and BML models.
These authors modeled streets with different capacities (i.e.,
different number of lanes) without
explicitly simulating several lanes.
They artificially reduced flux of streets with fewer than the maximum
number of lanes.
They applied their simulator to a fragment of the
Dallas/Fort-Worth area consisting of a 5-by-5-mile region.
The model considered route plans of 300,000 trips from
5:00 a.m.\ to 10:00 a.m.
For computational reasons, the computer program used random traffic
lights, which have the advantage of having to be checked only when a
vehicle reaches the intersection.

An approach closer to ours is that of Chopard, Luthi, and
Queloz~\citeyearpar{chopard:luthi:queloz:96}.
Each street has two lanes, each of which is bound for the opposite direction.
Traffic within streets follows rule 184.
At intersections, however, roundabouts mimic traffic lights as
follows.
On the one hand, vehicles within a roundabout have priority.
On the other hand, vehicles stay in the roundabout for a certain
number of ticks, after which all vehicles entering the roundabout
leave immediately for the same number of ticks.
This model exhibits metastability and gridlocks.
As a vehicle entering a roundabout must examine two cells so as to
determine whether or not it can actually enter the roundabout, this
cellular automaton is not elementary.

\section{An elementary model of city traffic}
\label{sec:model}

In this section, we extend
this model to develop a simple model of city traffic.
Rule 184 has been used as the simplest model of traffic flow
\citep{Yukawa:1994,ChowdhuryEtAl2000,Maerivoet:2005}.  We consider single lane streets that can go in four different directions, forming a Manhattan-style grid. This homogeneous setup is common of city traffic models \citep{FaietaHuberman1993,BrockfeldEtAl2001,Gershenson2005}.

The behavior of streets is already modeled with rule 184. If we draw
the cells horizontally, the vehicles will flow to the east. We can
obtain the other directions by rotating the street arrays by 90, 180,
and 270 degrees. Our extension lies in modeling intersections and
traffic lights. We can include them by considering several coupled non-homogeneous ECA, where rules change around the intersection, depending on the state of the traffic light.


Note that the combination of ECA has to be conservative \citep{Moreira:2003},
i.e.\ the number of 1's needs to be constant.
Otherwise, the model would be equivalent to having vehicles appearing or disappearing in the middle of the simulation.

If a street has a green light, all its cells use rule 184. If there is
a red light, all cells also use rule 184, with two exceptions: The
cell before (left of) the intersection has to stop traffic from going
into the intersection. Thus, if there is a vehicle in the actual cell
(010, 011, 110, 111), it will stay there, so the future state will
continue to be 1. If there is a vehicle in the previous cell (100,
101), it will advance, so the state will be 1. Only if there are no
vehicles in the previous and actual cells (000, 001), the state will
remain 0. This is rule 252. The cell after the intersection
has to allow vehicles to leave, but not to allow vehicles in the intersection (flowing in a different direction) to enter the cell. Thus, the future state will be 1 only if the cell ahead is blocked (011, 111). This is rule 136. Table \ref{table:ECArules} lists the transition tables for the three rules used by the model.

\begin{figure}[htbp]
\begin{center}
\includegraphics[width=15cm]{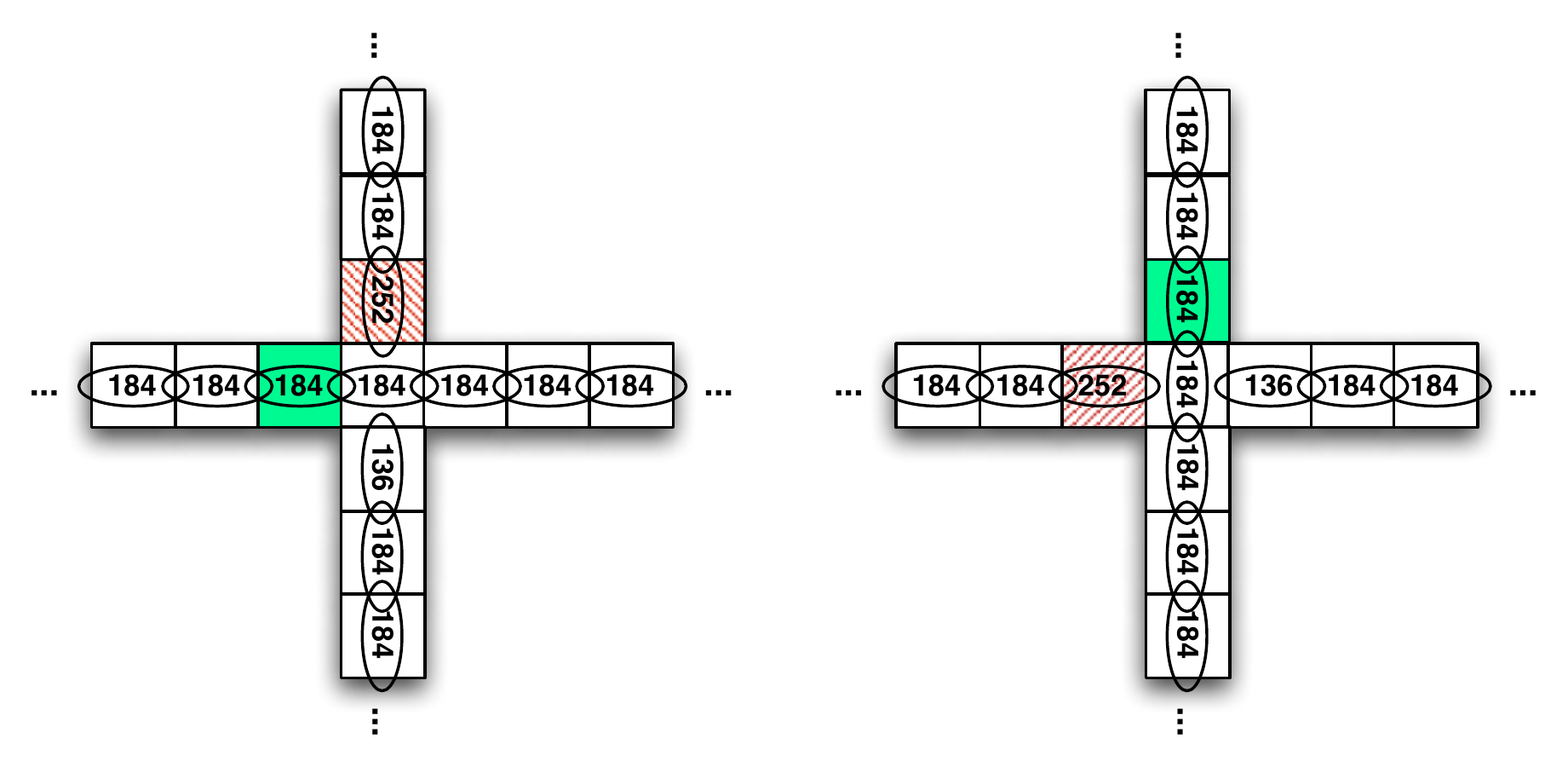}
\caption{Diagram for different rules (shown within cells) and neighborhoods (indicated by ovals) used around intersections, depending on the state of the traffic light. For green lights (indicated by a green solid cell), rule 184 is used, and the intersection cell has as neighbors cells in the street with the green light. For red lights (indicated by a diagonally red striped cell), rule 252 is used for the cell before the intersection and rule 136 for the cell after the intersection. The rest of the cells use rule 184.}
\label{rulesDiagram-single}
\end{center}
\end{figure}

The intersection cell is a special case, as it has four potential
neighbors. The rule never changes (184). What changes is the
neighborhood, i.e.\ it takes as nearest neighbors only the two cells in the street with a green light (also using rule 184). A diagram of the cells around an intersection is shown in Figure  \ref{rulesDiagram-single}.

\begin{table}[htdp]
\caption{ECA rules used in model}
\begin{center}
\begin{tabular}{|c|c|c|c|}
\hline
$t-1$	&$t_{184}$	&$t_{252}$	&$t_{136}$\\
\hline
000	&0	&0	&0	\\
\hline
001	&0	&0	&0	\\
\hline
010	&0	&1	&0	\\
\hline
011	&1	&1	&1	\\
\hline
100	&1	&1	&0	\\
\hline
101	&1	&1	&0	\\
\hline
110	&0	&1	&0	\\
\hline
111	&1	&1	&1	\\
\hline
\end{tabular}
\end{center}
\label{table:ECArules}
\end{table}%

If at time $t$ a traffic light is meant to switch, the model needs to
ensure that the intersection cell is empty. Otherwise, the vehicle in
the intersection would ``turn" into the crossing street. To avoid this
situation, the actual switching of rules and neighborhood is made only when the intersection is cleared.

Certainly, cells using rules 252 and 136 could be simplified to depend only on two cells, as their state is not affected by the intersection cell. However, we prefer to have redundancy in the rules but homogeneity in the neighborhood sizes.

To model flow in different directions, we could use mirror rules (e.g.\ 226 is equivalent to rule 184, with vehicles flowing to the left), but it is simpler to invert neighborhoods. Thus, the model is reduced to the combination of only three ECA rules.

\subsection{Measures}

The behavior of the model will depend strongly on the vehicle density
$\rho\in[0,1]$. Trivial cases are the extremes $\rho=0$, where there are no vehicles and $\rho=1$, where all cells are occupied by vehicles, so there is no space to move and flow is stopped. The density can be easily calculated by dividing the number of cells with 1 (i.e.\ total number of vehicles, $\sum{s_i}$) by the total number of cells ($\left|{S}\right|$):

\begin{equation}
\rho=\frac{\sum{s_i}}{\left|{S}\right|}
\label{eq:rho}
\end{equation}

The performance of the system can be measured with velocity $v\in[0,1]$, which is simply the number of cells that changed from 0 to 1 over the total number of vehicles:

\begin{equation}
v=\frac{\sum{{(s_i' > 0)}}}{\sum{s_i}}
\label{eq:v}
\end{equation}
where ${s_i'}$ is the derivative of state $s_i$. If $s_i'=1$, the cell
changed from 0 to 1. If $s_i'=-1$, the cell changed from 1 to
0. $s_i'=0$ when there is no change in the state of $s_i$,
i.e.\ there is no vehicle, or the vehicle at $s_i$ has stopped.

The flux of the system represents how much of the space is used by moving vehicles. It can be obtained by multiplying the vehicle density by the velocity:

\begin{equation}
J=\rho v
\label{eq:J}
\end{equation}

In the rule 184 model of highway traffic, the maximum possible flux is $J=0.5$, at a density $\rho=0.5$. This is because vehicles need at least one cell between them to move. If there are fewer vehicles, the flux will be lower, since there is no movement in free space. If there are more vehicles, then the flux will also be lower, since stopped vehicles do not move (See Figures \ref{fig:results_streetA} and \ref{fig:results_streetC}).

From equation \ref{eq:v}, the average waiting time and the number of stopped vehicles can be inferred as follows:

\begin{equation}
t_\I{wait} = \int{(1-v) dt}
\label{eq:wait}
\end{equation}

\begin{equation}
\I{vehicles}_\I{stopped} = (1-v)\sum{s_i}
\label{eq:stop}
\end{equation}

The average waiting time (\ref{eq:wait}) is calculated by integrating
over time the complement of the velocity ($1-v$). At every tick, if a
vehicle is not flowing, it means that such a vehicle is waiting, so $1-v$ captures how many vehicles are waiting.

The number of stopped vehicles (\ref{eq:stop}) is calculated by
counting all vehicles (1's) in the environment ($\sum{s_i}$) and
multiplying such a quantity by the complement of velocity ($1-v$), which indicates how many vehicles are not moving.

The percentage of stopped vehicles would be simply:

\begin{equation}
\%\I{vehicles}_\I{stopped} = 100(1-v)
\label{eq:Pstop}
\end{equation}

These measures can be calculated efficiently from the states of the automaton.

\subsection{Scales}

Even when the time and space are abstract and discrete, we can assume that one cell represents 5 meters, roughly the space occupied by a stopped vehicle. Thus, one kilometer of a street is represented by two hundred cells. If each tick, i.e.\ time step, represents one third of a second, then a velocity of one cell per tick is equivalent to 15 m/s, i.e.\ 54 km/h, roughly the speed limit in cities (e.g.\ in continental Europe it is 50 km/h in many countries; in Russia it is 60 km/h). A maximum density of $\rho=1$ is equivalent to 200 vehicles per kilometer.

\section{A single intersection}
\label{sec:single}

In this section, we treat a single intersection (we defer the case of a
grid, with many intersections, to Section \ref{sec:sims}).
We developed
a computer simulation in NetLogo \citep{Wilensky1999}. The reader is invited to access the simulation via web browser at the URL \url{http://turing.iimas.unam.mx/~cgg/NetLogo/trafficCA.html} (for short, \url{http://tinyurl.com/trafficCA}).
The environment consists of two cyclic streets, i.e.\ with periodic boundaries: one eastbound and one southbound, each of a length of 160 cells, one of which is shared, i.e.\ the intersection. Thus, a maximum density $\rho=1$ implies 319 vehicles.

For the traffic light, a period $T=160$ ticks was used. This implies
that the eastbound street has a green light for 80 ticks and the
southbound street has a green light for the following 80 ticks. Thus,
if a single vehicle is on the simulation, if it encounters a red
light, it will stop until the light turns green. Afterwards, the vehicle will always flow, since the time required to go once around the torus is equal to the period $T$.

\begin{figure}
     \centering
     \subfigure{
          \label{fig:results_singleA}
          \includegraphics[width=.45\textwidth]{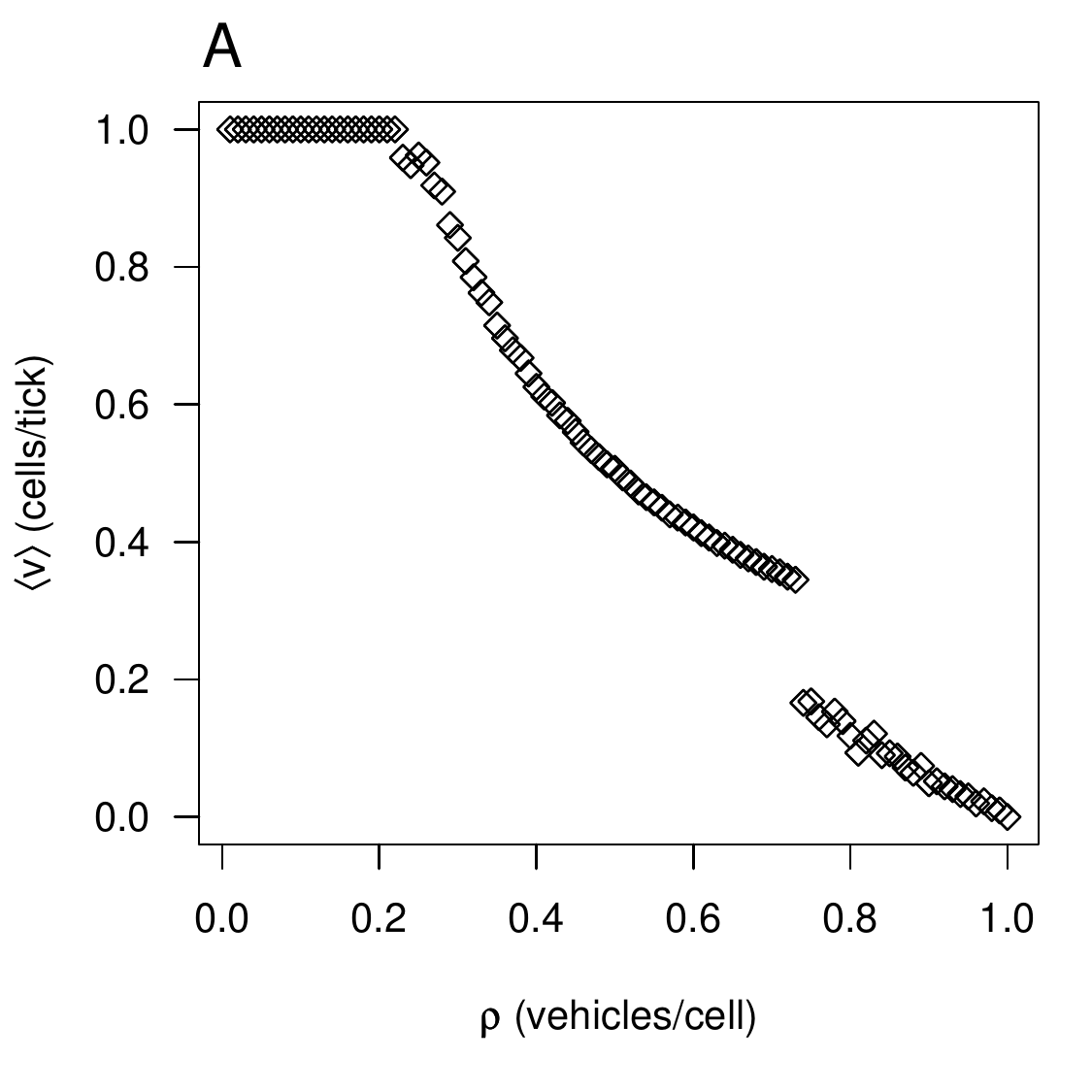}}
     \subfigure{
          \label{fig:results_singleB}
          \includegraphics[width=.45\textwidth]{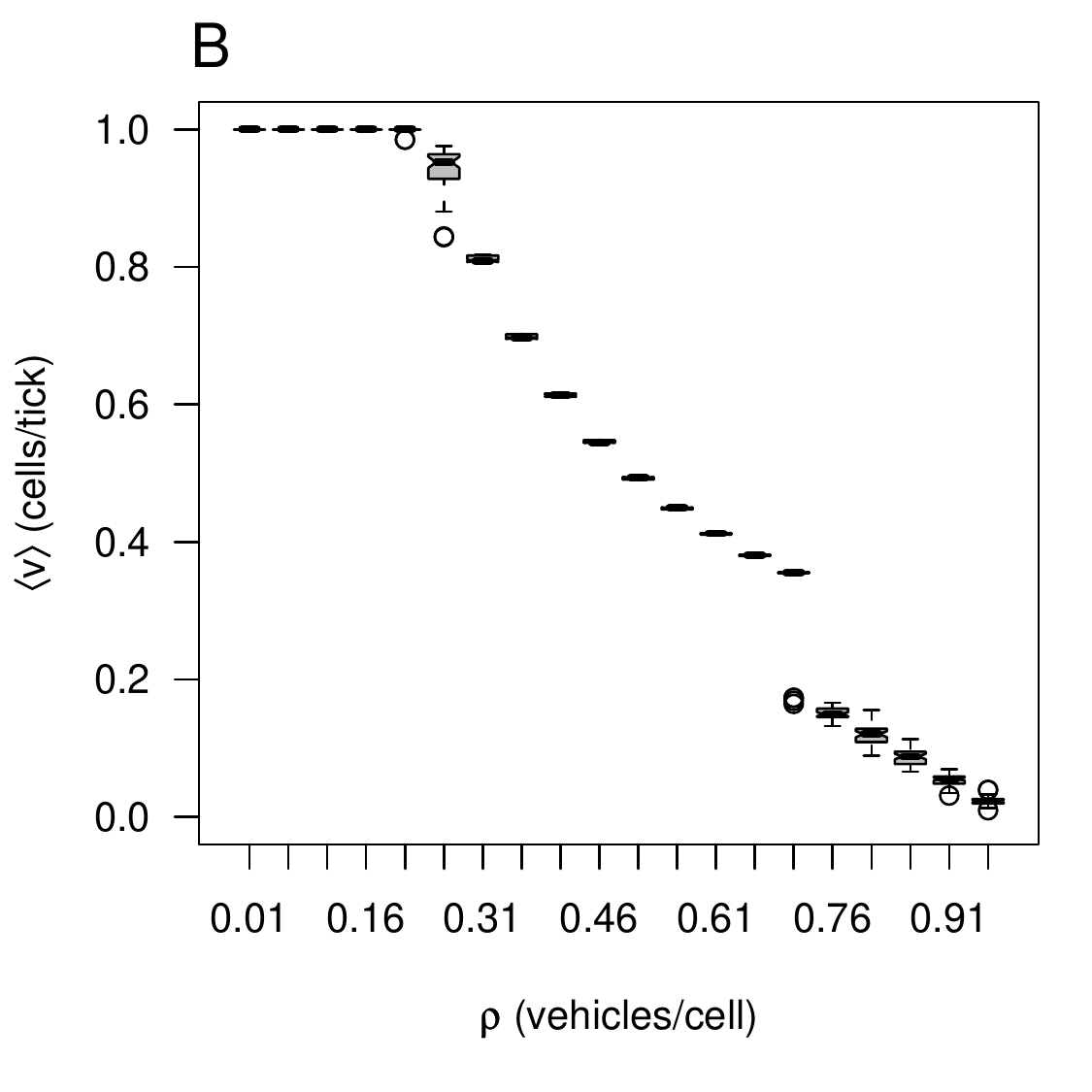}}
          \\
     \subfigure{
          \label{fig:results_singleC}
          \includegraphics[width=.45\textwidth]{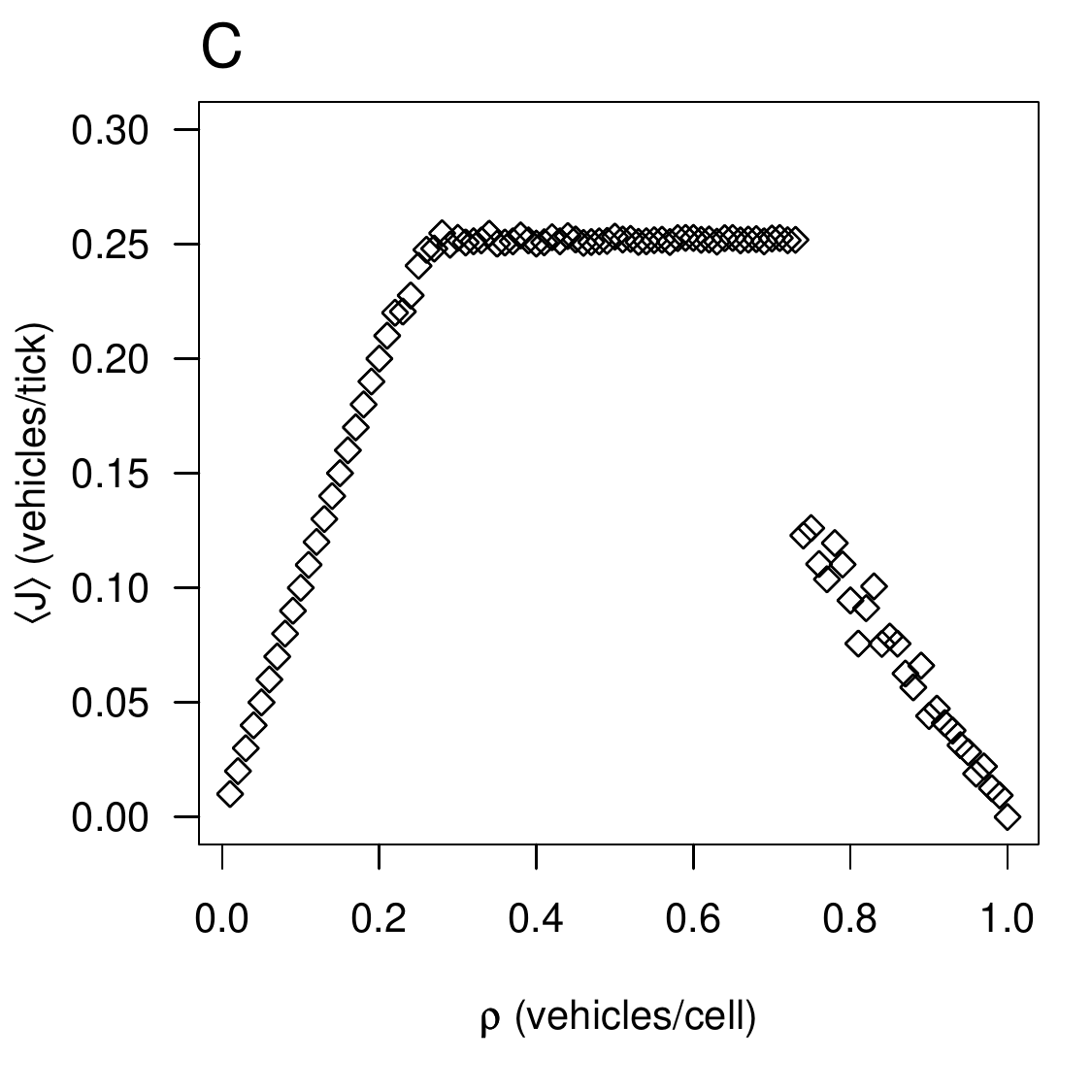}}
     \subfigure{
          \label{fig:results_singleD}
          \includegraphics[width=.45\textwidth]{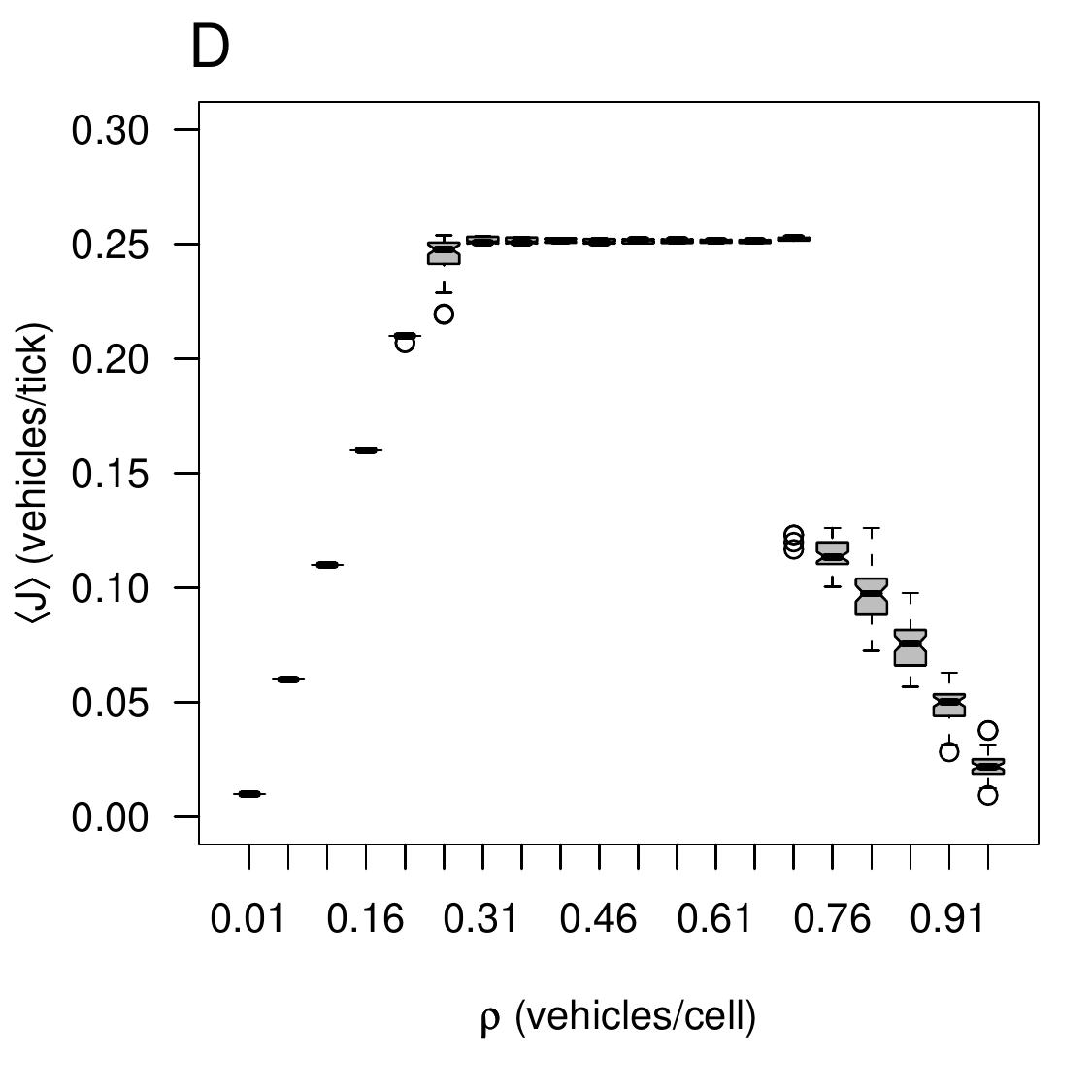}}

     \caption{Simulation results for a single intersection: (A,B) average velocity $\langle v\rangle$  and (C,D)  average flux $\langle J\rangle$ for different densities $\rho$: (A,C) single runs and (B,D) box plots of 50 runs per density.}
     \label{fig:results_single}
\end{figure}

For the experiments, each run consisted of the following: Half an hour (5,400 ticks) was simulated for random initial conditions. Since the vehicles are placed randomly, one street may have a slightly higher density than another. After these initial 5,400 ticks, the system is considered to have stabilized, i.e.\ gone through a transient, so another half an hour is simulated, of which velocity is measured at every tick. At the end of the simulation, the velocities of the second half an hour are averaged to obtain the average velocity $\langle v\rangle$ and average flux $\langle J\rangle$. The results are shown in Figure \ref{fig:results_single}.\footnote{A boxplot is a non-parametric representation of a statistical distribution. Each box contains the following information: The median ($Q2=x_{0.50}$) is represented by the horizontal line inside the box. The lower edge of the box represents the lower quartile ($Q1=x_{0.25}$) and the upper edge represents the upper quartile ($Q3=x_{0.75}$). The interquartile range ($IQR=x_{0.75}-x_{0.25}$) is represented by the height of the box. Data which is lesser than $Q1 - 1.5\cdot IQR$ or greater than $Q3 + 1.5\cdot IQR$ is considered an ``outlier", and is indicated with circles. The ``whiskers" (horizontal lines connected to the box) show the smallest and largest values that are not outliers.}

We can see that the phase transition from \emph{free-flow} ($v=1$) to an ``\emph{intermittent}" phase occurs at $\rho=0.25$. Recall that for no intersection, i.e.\ rule 184 as model of freeway traffic, a similar transition (to ``jammed" traffic) occurs at $\rho=0.5$, i.e.\ when there is exactly one free cell between vehicles. However, when an intersection is added, vehicles coming from both streets pass through the intersection cell. They have to share this resource, reducing the maximum flux to $J=0.25$. Thus, in order to have a free-flowing traffic (apart from setting $T$ carefully), there should be space left in one street while vehicles in the other are crossing. Otherwise, they have to stop behind a red light, leading to intermittent traffic. Notice that the average velocity and flux is reduced slightly before the phase transition at $\rho=0.25$. This is because there is a certain probability that one street will have a density $\rho>0.25$. Thus, not all of the vehicles on that street will be able to cross the intersection in one period and will have to wait, while the other street will have free-flow.

In this single-intersection model, there is a second phase transition at $\rho=0.75$ towards an ``\emph{interfered}" phase. This occurs when the traffic jams (traveling in the opposite direction of traffic at the velocity of one cell per tick) are long enough to reach the intersection around the torus and block it momentarily. This interference affects vehicles in the crossing street, reducing noticeably the average flux, i.e.\ $J=0.25$, before the phase transition and $J<0.125$ after it. The difference lies in the fact that when the interference of one street is dissolved, that same street will have the green light again, so in practice one street will be completely stopped. This explains why the flux $J$ is reduced to one half.

In the \emph{intermittent} phase ($0.25<\rho<0.75$), traffic jams form behind red lights and travel in the opposite direction of vehicles, but they dissolve before they reach again the intersection around the torus. This phase is characterized by a flux equal to the maximum of the model $J=0.25$, i.e.\ there are always vehicles crossing the intersection.

Since there is always some free space if $\rho<1$, the \emph{gridlock} situation ($v=0$) is only reached when $\rho=1$.

\begin{figure}
     \centering
     \subfigure{
          \label{fig:results_single_TA}
          \includegraphics[width=.45\textwidth]{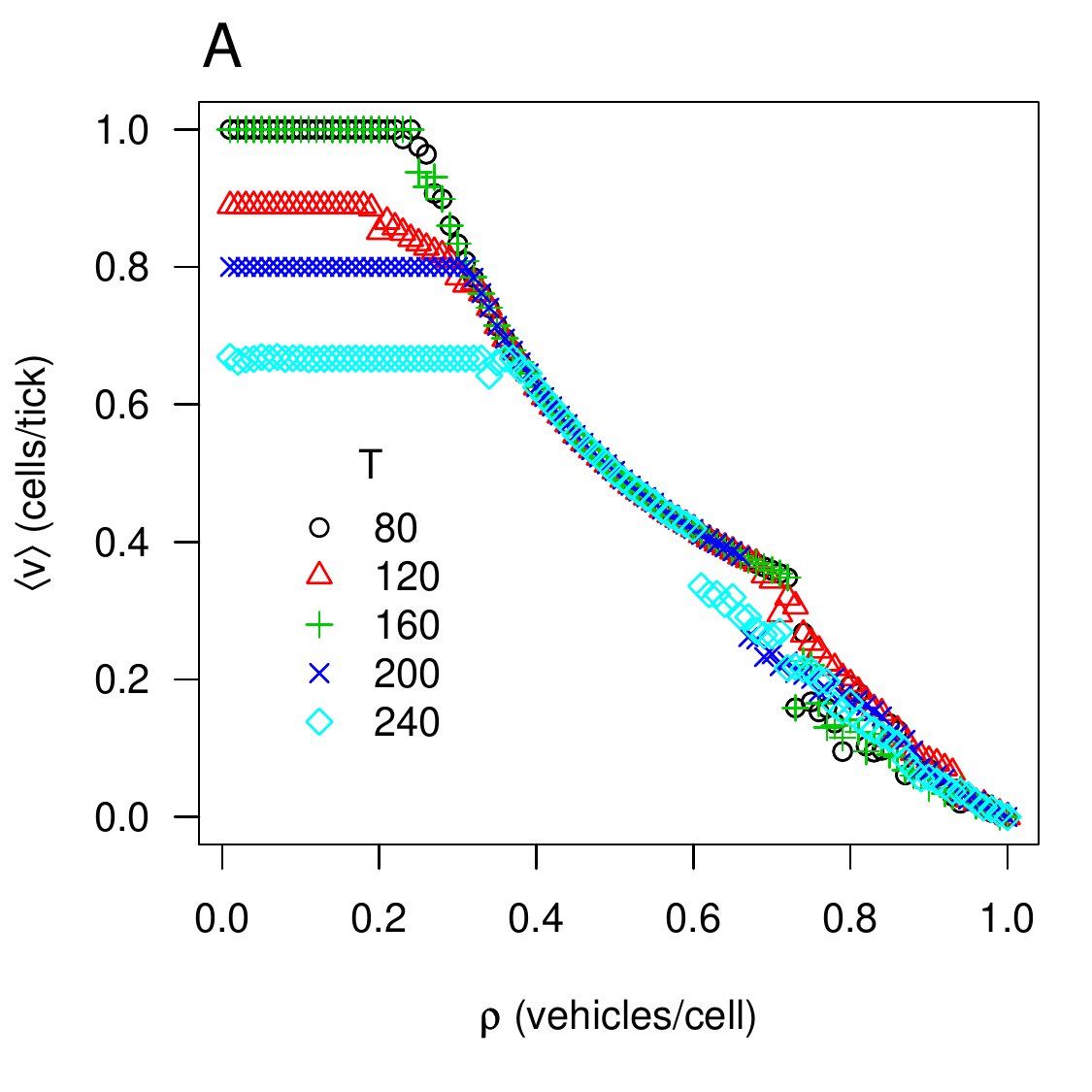}}
     \subfigure{
          \label{fig:results_single_TB}
          \includegraphics[width=.45\textwidth]{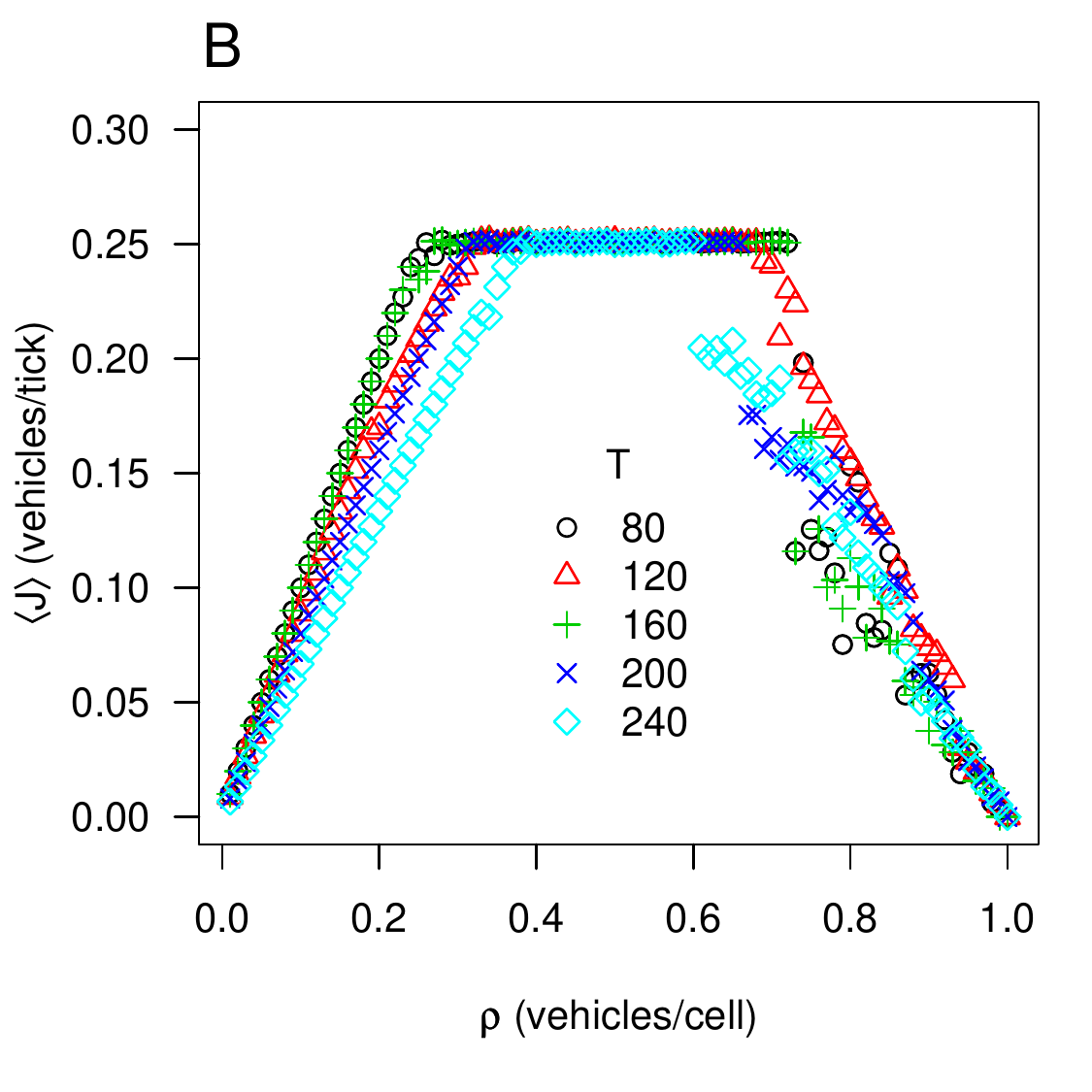}}

     \caption{Simulation results for a single intersection varying period $T$. (A) Average velocity $\langle v\rangle$   and (B) average flux $\langle J\rangle$  for different densities $\rho$.}
     \label{fig:results_single_T}
\end{figure}

We performed experiments where $T$ was varied, shown in Figure
\ref{fig:results_single_T}. We can see in Figure \ref{fig:results_single_TA} that the value of $T$ affects
the existence of a \emph{free-flow} phase. This requires a synchronization of the traffic
light period $T$ with the travel time around the torus. This is
achieved in our model only the street length is a multiple of 
$T$, a condition for free-flow. Otherwise, we can say that the
period of the traffic light and the period of the vehicles are out of
synchrony. Thus, vehicles need to wait for a green light and the average
velocity is less than one. However, once the density reaches the
\emph{intermittent}, phase, the period $T$ does not affect the performance of the system, i.e.\ there is always a maximum flux (see Figure \ref{fig:results_single_TB}). Still, there is a certain symmetry in the flux diagram, in the sense that the $T$ values that reach the maximum flux capacity earlier will degrade later, i.e.\ values that reach the \emph{intermittent} phase earlier will reach the \emph{interfered} phase later. However, within the \emph{interferred} phase, the synchronization of $T$ with vehicular travel time is counterproductive, because in most cases one street will be blocked (see discussion above). The other $T$ values give a better performance because---even when flow is \emph{interferred}---all vehicles are able to move. Thus, there is no single $T$ value that gives the best performance across all densities $\rho$.

\section{Methods for coordinating traffic lights}

We present in this section two methods for controlling traffic lights.
Given the interactions and feedbacks among flows exiting and entering intersections, it is not possible to generalize the results of a single intersection to a city grid.
It is known that the optimal coordination of traffic lights is an EXP-complete problem \citep{PapadimitriouTsitsiklis1999,Lammer:2008}. This implies that it is intractable. There have been several methods proposed over recent decades to solve this problem. We can distinguish two main approaches. One tries to \emph{optimize} the green phases of lights to maximize the movement of vehicles traveling at a certain velocity \citep{FHA2005,Robertson:1969,N.H.Gartner:1975,SCATS1980,TorokKertesz1999,BrockfeldEtAl2001}. The other one tries to \emph{adapt}---manually or automatically---phases depending on current traffic flows \citep{FHA2005,Henry:1983,Mauro:1990,Robertson:1991,FaietaHuberman1993,Gartner:2001,Diakaki:2003,FouladvandEtAl2004a,Mirchandani:2005,Bazzan2005,HelbingEtAl2005,Gershenson2005,CoolsEtAl2007}. We implemented two methods, one corresponding to each approach: a \emph{green-wave} method that tries to optimize phases for an expected traffic flow, and a \emph{self-organizing} method that adapts to the current traffic conditions.

\subsection{The \emph{green-wave} method}
\label{sec:gw}

The idea behind the \emph{green-wave} method is the following: if the
consecutive traffic lights switch with an offset (i.e.\ delay) equivalent to the expected vehicle travel time between intersections, vehicles should not have to stop. Thus, waves of green light move through the street at the same velocity as vehicles.

This method has advantages, e.g.\ when most of the traffic flows in
the direction of the green wave at low densities. However, since only two directions
can have green waves, vehicles flowing in the opposite direction of
the green wave will be delayed. Also, if traffic is flowing at
velocities lower than expected, the green waves will go faster than vehicles and these will be delayed.

To implement the \emph{green-wave} method in our CA model, we consider a fixed period $T$ for all traffic lights, and an individual offset $\omega_j$ for each traffic light. This offset is determined by the coordinates $x,y$ of the intersection cells as follows:

\begin{equation}
\omega_j = \lfloor(( x - y) \bmod \frac{T}{2}) + 0.5 \rfloor
\label{eq:offset}
\end{equation}

Equation \ref{eq:offset} sets the offset by rounding to the nearest integer the $x$ coordinate minus the $y$ coordinate, modulus half a period. This offset creates green waves to the south and east, for regular or irregular grids. Lights are switched twice per period $T$, when $\omega_i$ is equal to a global phase $\varphi$ that cycles from zero to $\frac{T}{2}$.
To allow vehicles to flow without stopping, the street lengths need to be a multiple of period $T$, just as with the single intersection case. The main idea behind setting the offset is to have intersections lying on the skew diagonals to have the same offset value, so as to switch their lights at the same time. In this way, free-flowing vehicles going either eastbound or southbound will be able to reach an intersection with a green light. To change the direction of the green wave, only the sign of the $x$ or $y$ coordinates has to be changed.

The lights need to be initialized properly for the \emph{green-wave} method to work. They can be in a state of a green light horizontally (east or westbound) and thus red light vertically (north or southbound), or a state of a green light vertically and thus a red light horizontally.
The state $\sigma_{j}$ of the traffic lights is initialized as follows:

\begin{equation}
\sigma_{j} = \left \{ \begin{matrix}
\I{green}_\I{vertical} & \mbox{if}& \lfloor((x - y)\bmod T)+0.5 \rfloor \geq \frac{T}{2}
\\
\I{green}_\I{horizontal}& \mbox{otherwise}
\end{matrix}\right.
\label{eq:init}
\end{equation}

Equation \ref{eq:init} initializes the state of a traffic light to have a green vertical light  if the nearest integer of the $x$ coordinate minus the $y$ coordinate, modulus one period $T$, is greater than or equal to half a period. Otherwise, the state is set as green horizontal. With this equation, 50\% of contiguous intersections of every street (vertical  and horizontal) will have a green vertical state and the other half a green horizontal state. The states of intersections are arranged in such a way that the transitions from green horizontal to green vertical lie on skew diagonals.

\subsection{The \emph{self-organizing} method}
\label{sec:so}

With the \emph{self-organizing} method, each intersection
independently follows the same set of rules, based only on local traffic information. There are only six rules (not related to ECA rules), with higher-numbered rules overriding lower-numbered ones. The full rule set is given in Table \ref{table:rules}. This method is an improvement over previous work. The one reported in \citep{Ball:2004} considered only rules 1 and 2, while \citep{Gershenson2005,CoolsEtAl2007}, considered only rules 1--3.

\begin{table}[htdp]
\caption{\emph{Self-organizing} traffic light rules. Inset: Schematic of an intersection, indicating distances $d$, $r$, and $e$ used for self-organizing lights.}
\begin{center}
\vspace{0.2cm}
\fbox{
\parbox{6 in}{

 \begin{center}
   \includegraphics[height=30mm]{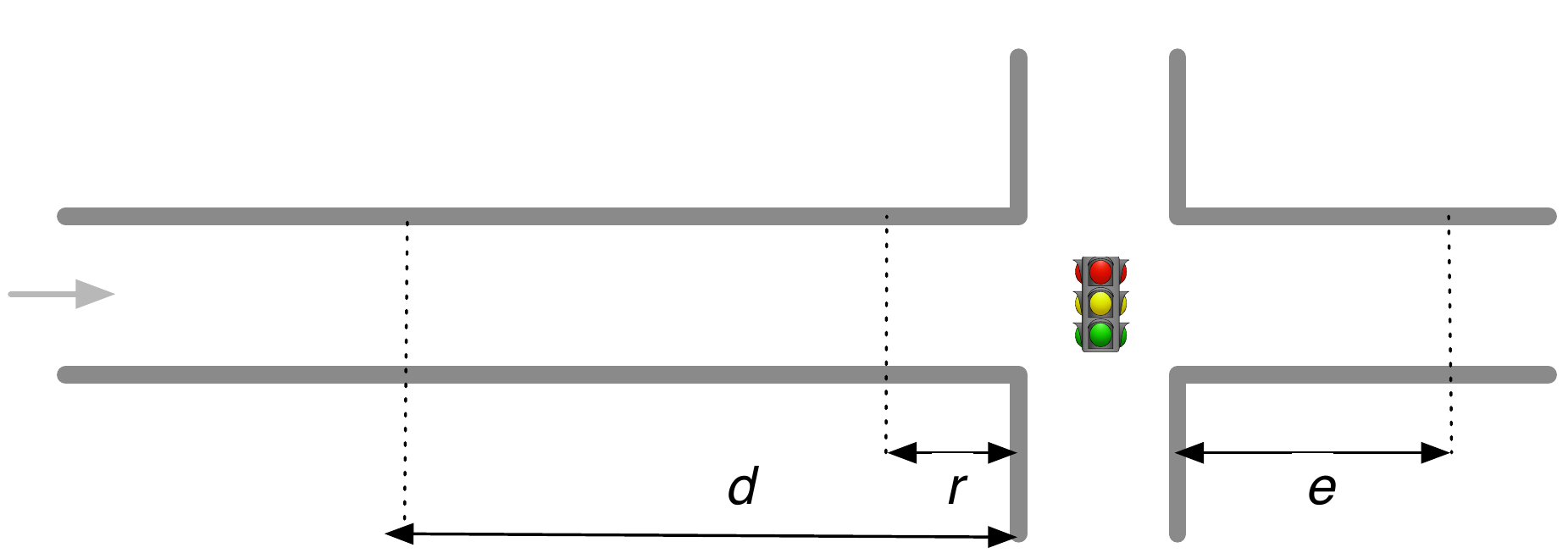}
 \end{center}

{\small
\begin{enumerate}
\item On every tick, add to a counter the number of vehicles
approaching or waiting at a red light
within distance $d$.
When this counter exceeds a threshold $n$, switch the light.  (Whenever the light switches, reset the counter to 0.)
\item Lights must remain green for a minimum time $u$.
\item If a few vehicles ($m$ or fewer, but more than zero) are left to cross a green light at a short distance $r$, do not switch the light.
\item If no vehicle is approaching a green light within a distance $d$, and at least one vehicle is approaching
the red light within a distance $d$, then switch the light.
\item If there is a vehicle stopped on the road a short distance $e$
  beyond a green traffic light, then switch the light.
\item If there are vehicles stopped on both directions at a short distance $e$
  beyond the intersection, then switch both lights to red. Once one of the directions is free, restore the green light in that direction.
\end{enumerate}

}

}

}

\end{center}
\label{table:rules}
\end{table}%

The first rule is designed to ensure that when traffic is waiting at a red light, or many vehicles are approaching an intersection,
the light will switch to green. Since we are concerned about the idle
time at intersections, a possible quantity to use for triggering
lights is the accumulated waiting time of vehicles. However, it is even
better to include the vehicles approaching a light, so they may not
need to slow down or stop at all. Thus, every light has a counter that records the cumulative amount of vehicle time within a set distance
$d$ from the light since it last changed to red, adding to this value every tick.
All incoming vehicles within distance $d$ are counted, whether stationary or moving.
When the counter exceeds a threshold $n$, the light is switched (subject to override by subsequent
rules). Thus, for example, if one vehicle waits for 40 ticks, 5 vehicles wait for 8 ticks, or 10 vehicles wait for 4 ticks, and then the light will be ready to switch. If there are many incoming vehicles approaching a red light, rule 1 will tend to switch their light to green before they
reach the intersection, so they will not need to stop. Vehicles waiting at an intersection may be joined by others to form
a platoon before the light switches.
As the platoon flows through the system, its approach switches other lights to green, creating an emergent green wave.

Rule 2 prevents platoons approaching the same intersection from conflicting directions from triggering repeated switching that would immobilize traffic.
Such a rule sets a minimum time before a platoon can request a light change.
The integrity of platoons is promoted by rule 3, which prevents the ``tails'' of platoons from being cut,
but allows the division of long platoons. Rule 4 allows rapid
switching of lights for low traffic densities, so lone vehicles can trigger lights to switch as they approach
an intersection without needing to wait for platoons to be formed. Rules 5 and 6 prevent gridlock otherwise caused by halting vehicles before they might block an intersection due to stopped vehicles on the other side of the intersection. Rule 5 changes the light if there is a blockage ahead of a green light, while rule 6 sets both lights to red if both directions are blocked.
A formal description of the \emph{self-organizing} method is presented in Appendix \ref{App:Algo}.

A variation of rule 1 has been in used in the U.K.\ for more than twenty years \citep{VincentYoung1986}, but only at
a limited number of isolated intersections. The technology to implement this method has already been on the market for several years, improving constantly the sensor accuracy and sophistication.

\section{Simulations of city traffic}
\label{sec:sims}

To test the model and to compare the above-mentioned traffic light controllers, we used again a street distance of 160 cells, i.e.\ 800 meters, but now we constructed a ten-by-ten homogeneous grid, with alternating flow directions and cyclic boundaries. Thus, there are 80 simulated meters between streets.\footnote{The interstreet distance in Manhattan is 75 meters. Flow directions also alternate.} There are 3,100 cells in the simulation, one hundred intersections, and twenty streets: five in each cardinal direction.

\begin{figure}
     \centering
     \subfigure{
          \label{fig:results_cityA}
          \includegraphics[width=.45\textwidth]{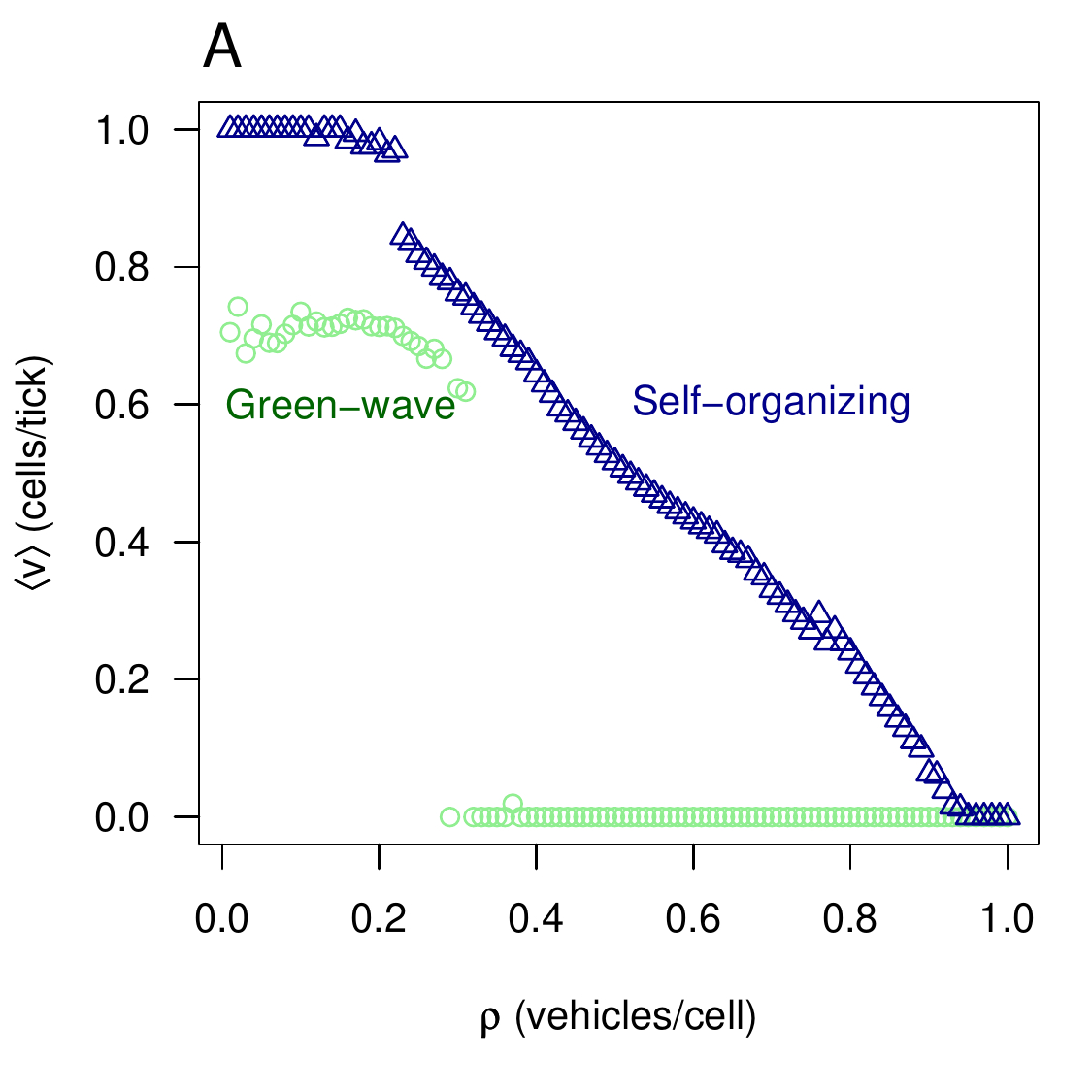}}
     \subfigure{
          \label{fig:results_cityB}
          \includegraphics[width=.45\textwidth]{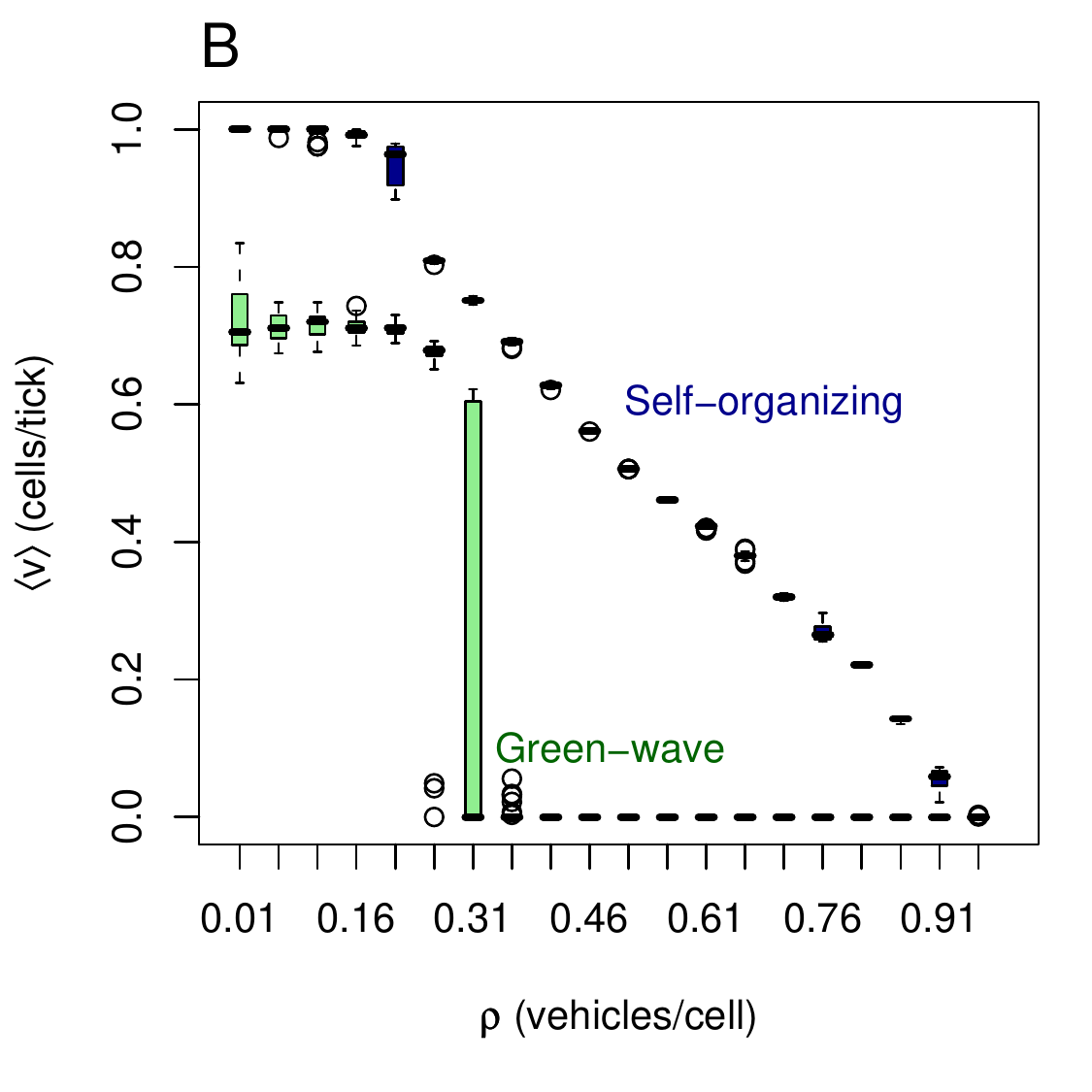}}
     \\     
     \subfigure{
          \label{fig:results_cityC}
          \includegraphics[width=.45\textwidth]{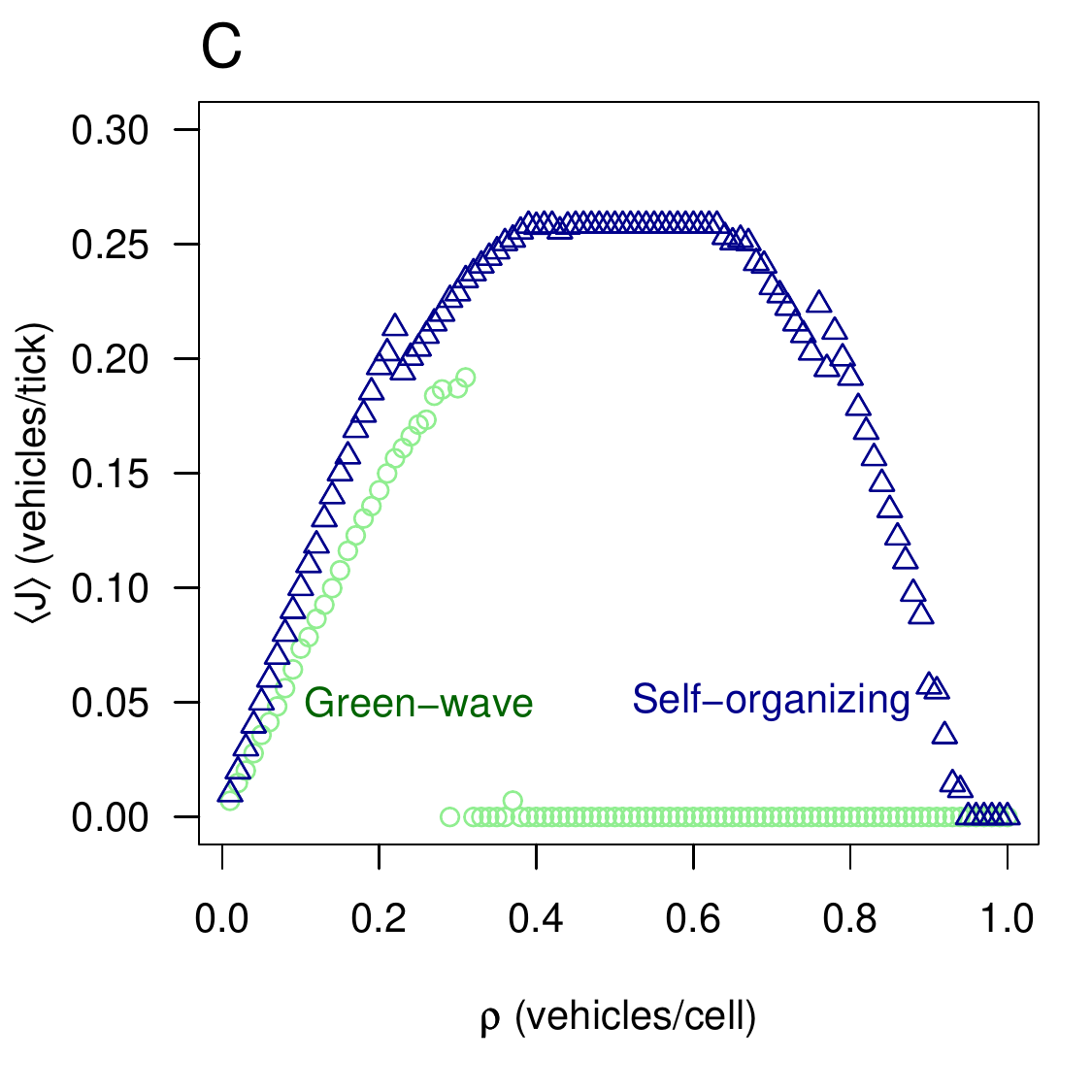}}
     \subfigure{
          \label{fig:results_cityD}
          \includegraphics[width=.45\textwidth]{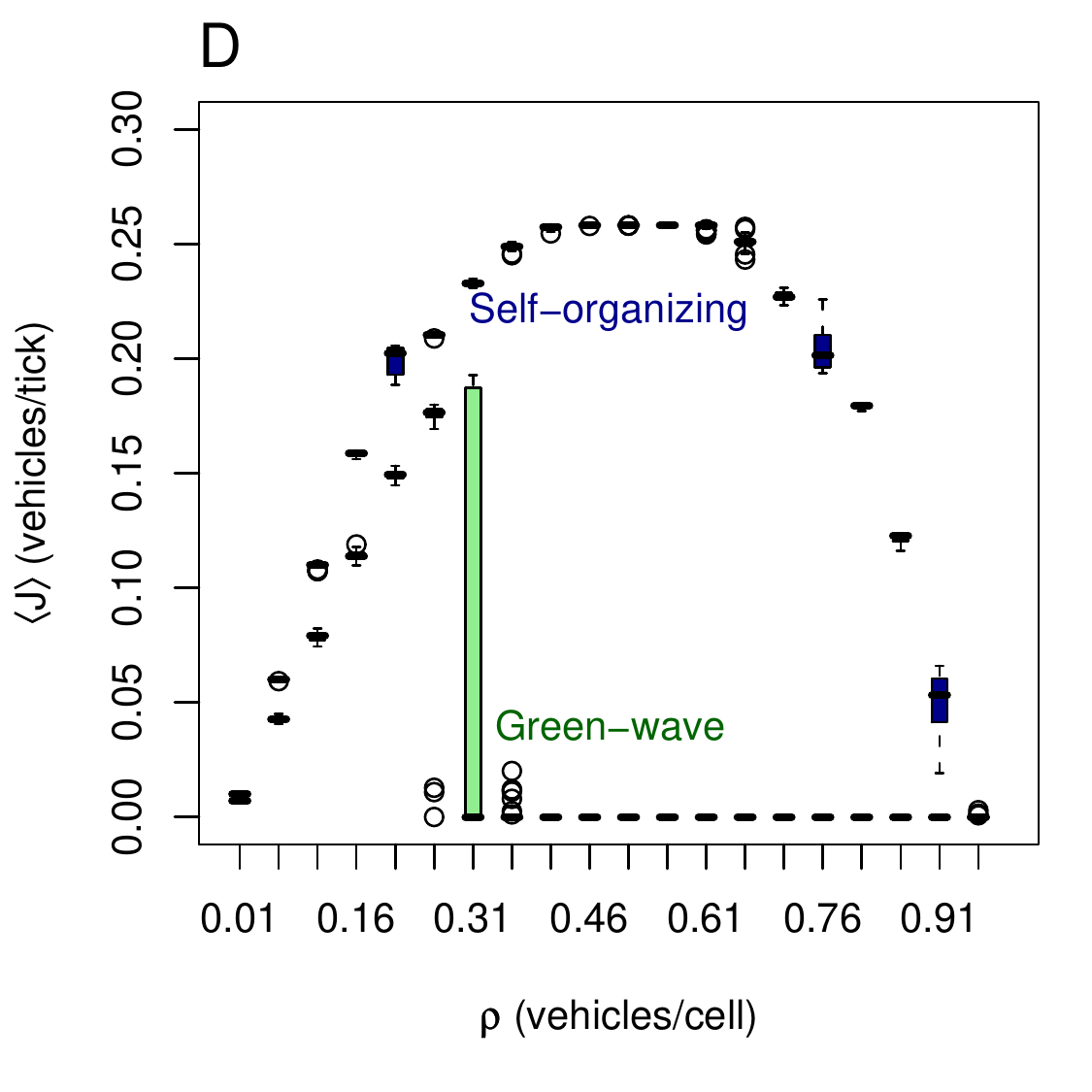}}

     \caption{Simulation results for a ten-by-ten city grid: (A,B) average velocity $\langle v\rangle$ and (C,D) average flux $\langle J\rangle$ for different densities $\rho$, \emph{green-wave} ({\Large $\circ$}) and self-organizing ($\bigtriangleup$) methods: (A,C) single runs and (B,D) box plots of 50 runs per density.}
     \label{fig:results_city}
\end{figure}

We used the same simulation developed in NetLogo and experimental setup as the one for the single intersection experiments, described in Section \ref{sec:single}. The simulation is available at the URL \url{http://tinyurl.com/trafficCA}. Results are shown in Figure \ref{fig:results_city}.

\begin{figure}
     \centering
     \subfigure[]{
          \label{fig:phases_gwA}
          \includegraphics[width=.45\textwidth]{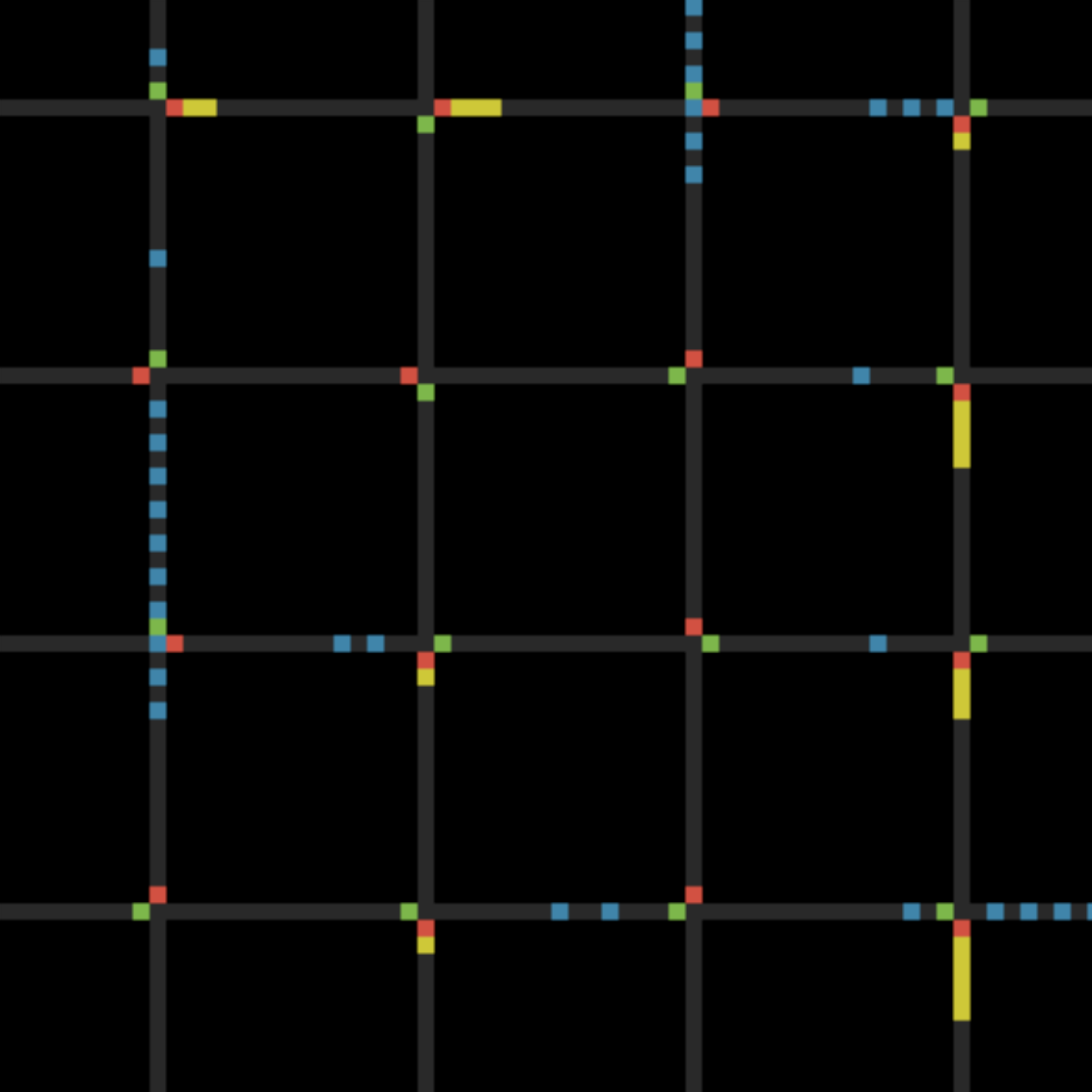}}
     \subfigure[]{
          \label{fig:phases_gwB}
          \includegraphics[width=.45\textwidth]{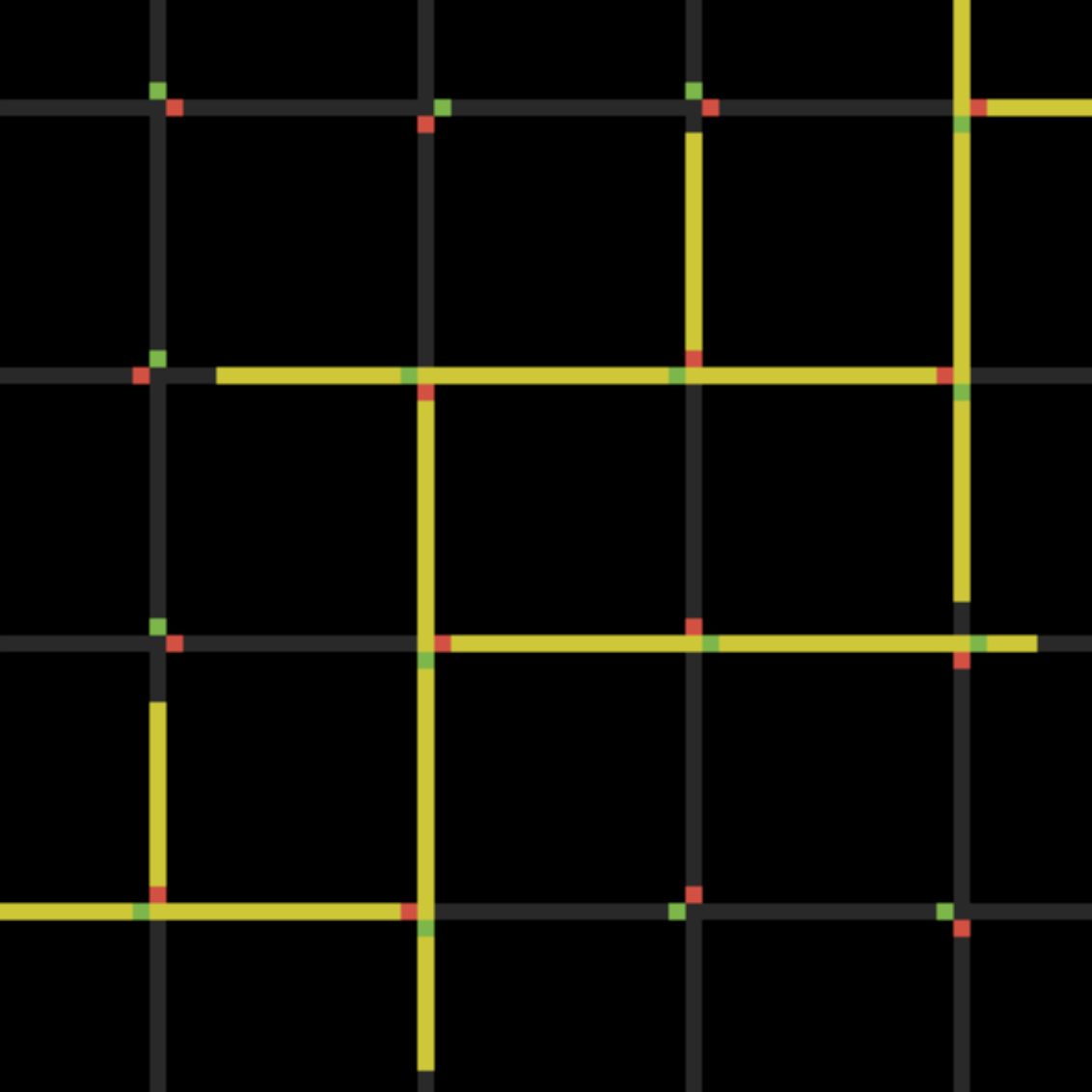}}

     \caption{Screenshots of different phases for \emph{green-wave} method. Four-by-four sections of ten-by-ten simulations shown. Blue cells indicate moving vehicles, yellow cells indicate stopped vehicles. (A) \emph{Intermittent} phase ($\rho = 0.1$): Vehicles flow freely on streets with green waves, but vehicles going in the opposite direction have to stop frequently. (B) \emph{Gridlock} phase ($\rho = 0.3$): Intersection blockages propagate and all vehicles stop, even when the density is not high.}
     \label{fig:phases_gw}
\end{figure}

For the \emph{green-wave} method, at densities $\rho < 0.25$ half of the streets (southbound and eastbound) have a free-flow regime, i.e.\ $v=1$. This is comparable with the behavior of the single intersection at these densities. However, this means that on the other half of the streets (northbound and westbound), the velocity is half the one shown in Figure \ref{fig:results_city} ($v \approx 0.35$). In these streets, vehicles stop every three blocks, leading to large delays due to the anti-correlated traffic lights they encounter. We found only two phases: an \emph{intermittent} phase, where some vehicles have to stop at traffic lights and a \emph{gridlock} phase, i.e.\ $v=0$. The phase transition lies at $\rho \approx 0.3$, where there is a maximum flux of $J \approx  0.19$. For densities $0.25 < \rho \lessapprox 0.3$, vehicles flowing on streets with green wave (southbound and eastbound) cannot keep the speed of the green wave, so traffic jams from that move in the opposite direction of the traffic.
At densities $\rho \gtrapprox 0.3$, the queues on directions opposite to the green wave grow and block intersections upstream, leading to gridlocks at medium and high densities. In Figures \ref{fig:results_cityB} and \ref{fig:results_cityD} there is a high variance around the phase transition because the gridlock outcome depends on which percentage of the randomly placed vehicles lie on streets with anti-correlated lights. Screenshots of the phases for the \emph{green-wave} method can be seen in Figure \ref{fig:phases_gw}.

For the \emph{self-organizing} method, five phases were found, one of which had three subphases:
\begin{enumerate}
\item For low densities there is a \emph{free-flow} phase ($v=1$) and no vehicle has to stop. 
\item There is a phase transition at $\rho \approx 0.15$ into a \emph{quasi-free-flow} phase, where very few vehicles stop for very little time. Most intersections have only one platoon requesting a green light, so this one is able to flow without having to wait for other platoons.
\item At $\rho \approx 0.22$ there is another phase transition into an \emph{intermittent} phase, where few or some vehicles stop behind traffic lights. 
Within the intermittent phase, we can distinguish three subphases depending on the flux diagram (Figure \ref{fig:results_cityC}): 
\begin{enumerate}
\item Until $\rho \approx 0.38$, there is an \emph{underutilized intermittent} subphase, where intersections are idling some of the time, i.e.\ no vehicle uses them. The difference with the \emph{quasi-free-flow} phase is that in the \emph{underutilized intermittent} subphase there are two platoons requesting a green light in most of the cases. Thus, one platoon has to wait until the other one crosses.
\item Between $\rho \approx 0.38$ and $\rho \approx 0.63$ there is a \emph{full capacity intermittent} subphase, where there is a maximum flux $J=0.25$. This implies that the intersections are being used at their maximum capacity, i.e.\ there is a vehicle crossing every other tick, so there are no ``wasted" resources of free space. Notice that this subphase occupies one fourth of the density space $\rho \in [0,1]$.
\item After $\rho \approx 0.63$ and until $\rho \approx 0.77$ there is an \emph{overutilized intermittent} subphase, where the density is such that rule 6 sometimes forces both directions to stop, thus reducing the flux of the intersections. This subphase is similar to the \emph{underutilized intermittent} subphase, in the sense that  the intersections cannot be used at their full flux capacity. In the \emph{underutilized intermittent} case, this is because there are not enough vehicles. In the \emph{overutilized intermittent} case, this is because there are too many vehicles and intersections need to wait before one street can get a green light.
\end{enumerate}
\item Between $\rho \approx 0.77$ and $\rho \approx 0.95$ there is a \emph{quasi-gridlock} phase. Most vehicles are stopped, but free spaces ``move" in the direction opposite of the vehicles between traffic jams at a speed of one cell per tick. Adaptive green waves also move in that direction, i.e.\ free spaces always trigger a green light. Just like the \emph{self-organizing} method promotes the formation of platoons and these coordinate the traffic lights using mainly rule 1 for low densities, the method promotes the formation of free spaces that coordinate traffic lights using mainly rule 6 for high densities. These free spaces allow vehicles to advance for longer distances before stopping. Their coordination implies that there will be little interference between free spaces traveling in different streets, i.e.\ free spaces rarely have to stop at intersections, and they do so only to allow other free spaces to finish crossing. The difference with the \emph{overutilized intermittent} phase is that in the \emph{quasi-gridlock} phase most traffic lights are switched by rule 6, while in the \emph{overutilized intermittent} phase other rules also play a role, i.e.\ free spaces are long enough to meet at intersections, so vehicles approaching intersections can trigger traffic lights.
\item There is a final transition at $\rho \approx 0.95$ into a gridlock phase, i.e.\ $v=0$. This is due to initial conditions, where some streets are blocked before the \emph{self-organizing} method can prevent this.

\end{enumerate}

\begin{figure}
     \centering
     \subfigure[]{
          \label{fig:phases_so1A}
          \includegraphics[width=.45\textwidth]{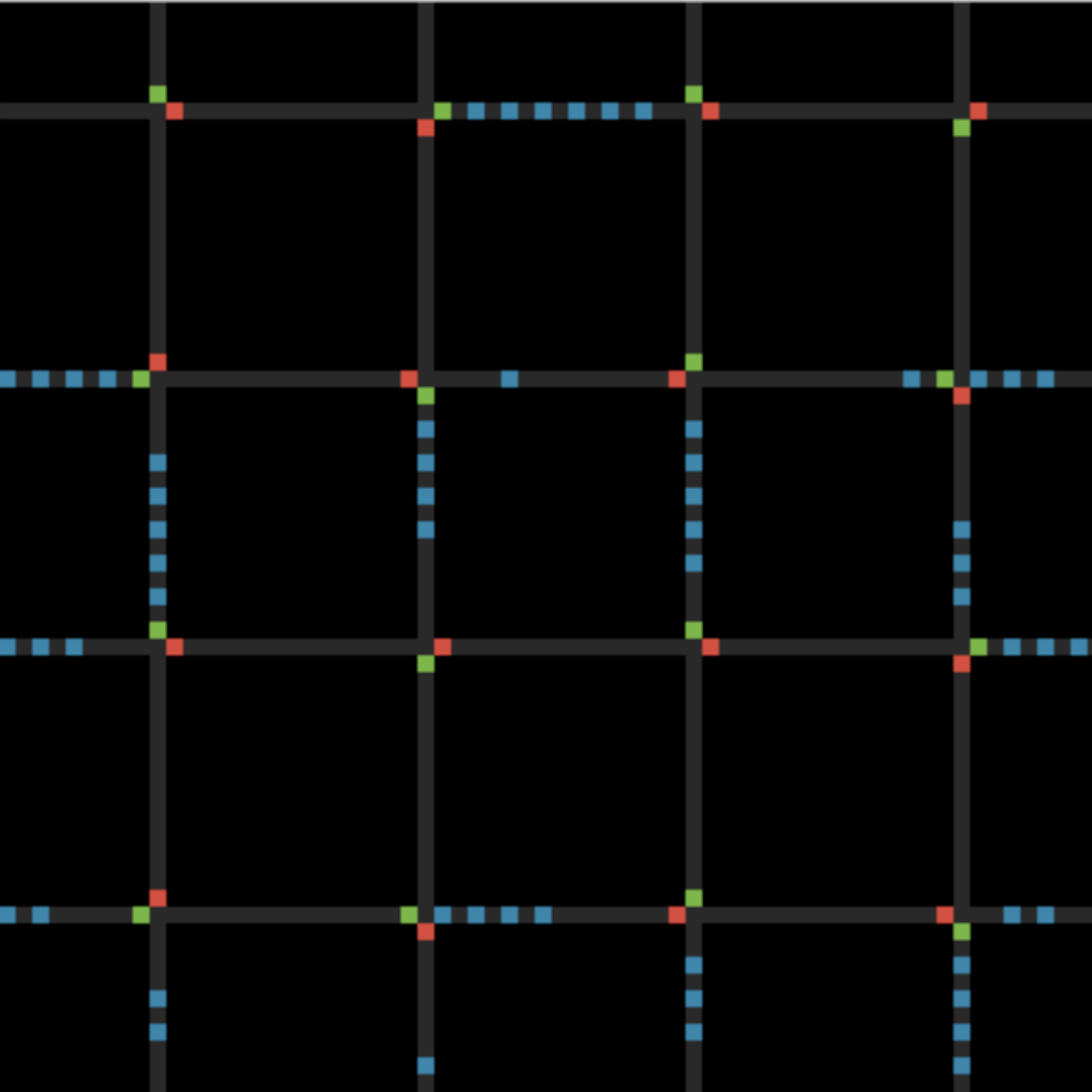}}
     \subfigure[]{
          \label{fig:phases_so1B}
          \includegraphics[width=.45\textwidth]{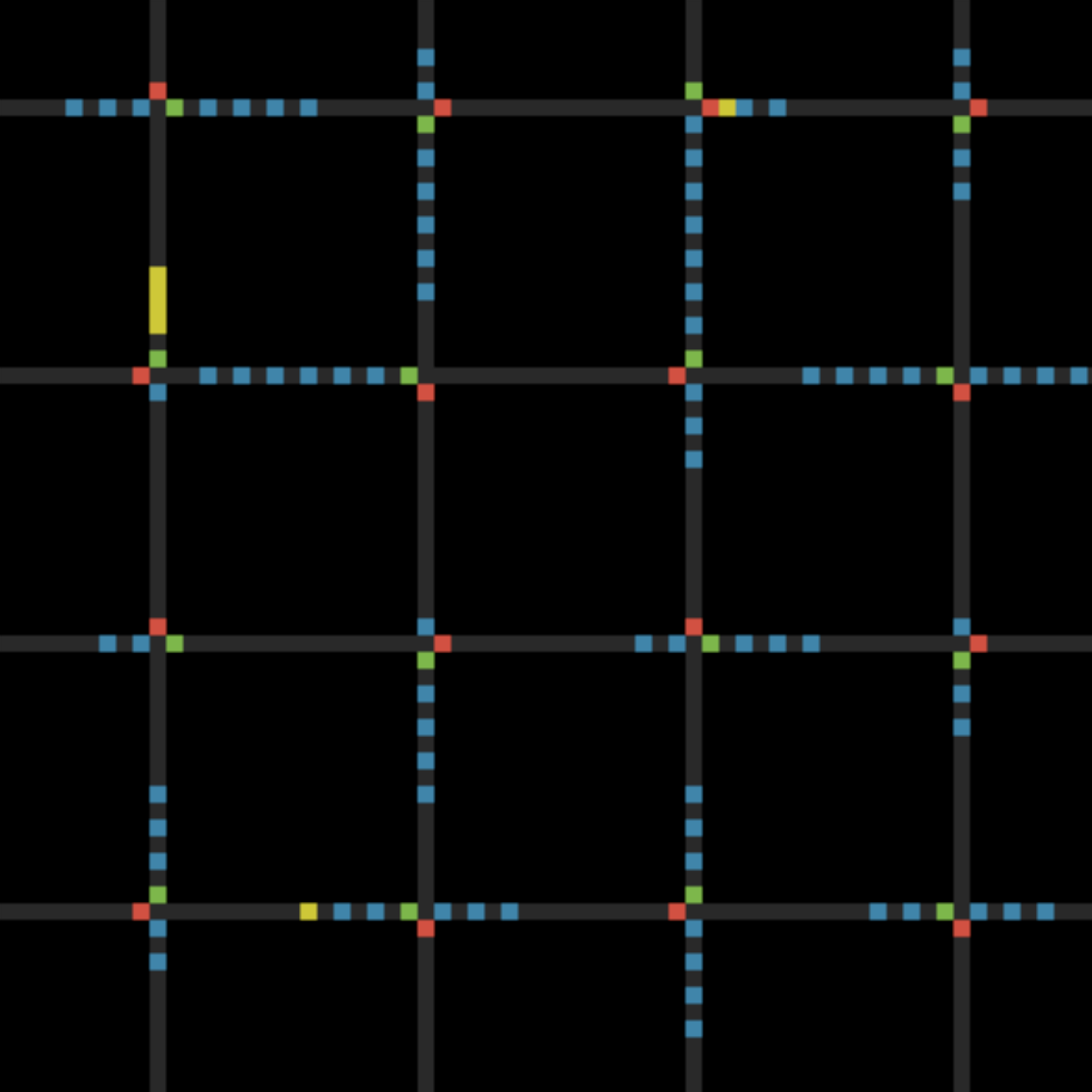}}
\\
     \subfigure[]{
          \label{fig:phases_so1C}
          \includegraphics[width=.45\textwidth]{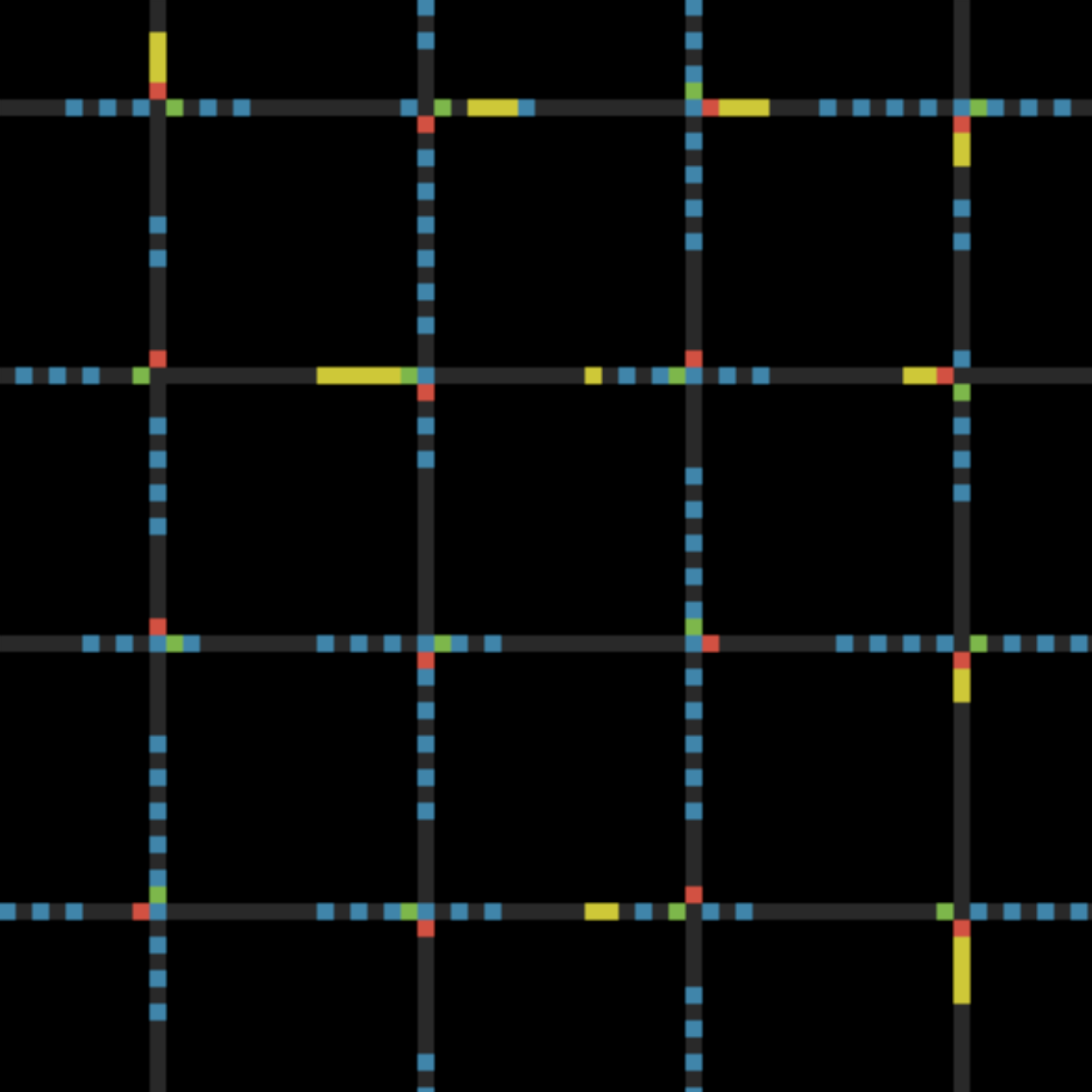}}
     \subfigure[]{
          \label{fig:phases_so1D}
          \includegraphics[width=.45\textwidth]{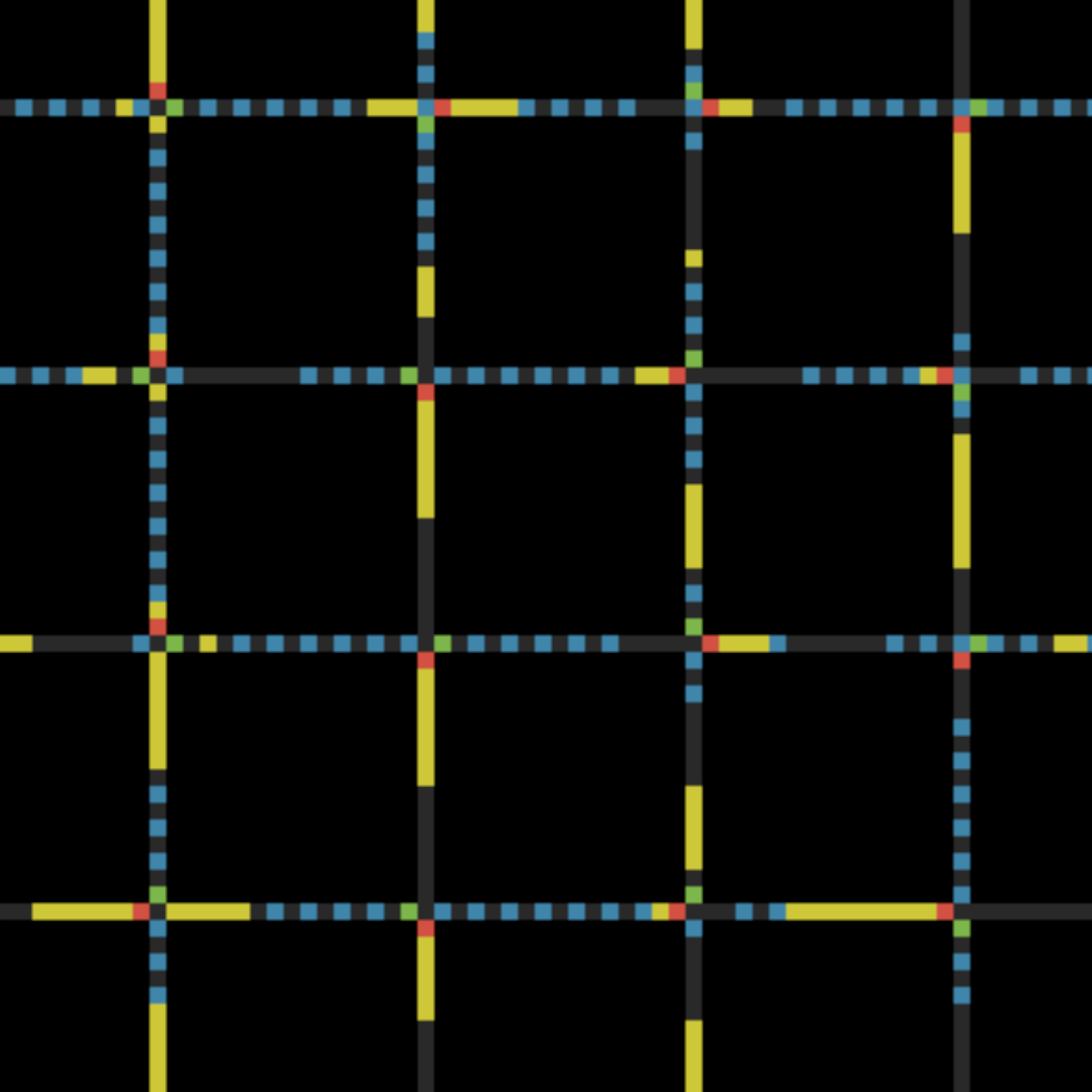}}

     \caption{Screenshots of different phases for \emph{self-organizing} method. Four-by-four sections of ten-by-ten simulations shown. Blue cells indicate moving vehicles, yellow cells indicate stopped vehicles. See also Figure \ref{fig:phases_so2}. (A) \emph{Free-flow} phase ($\rho = 0.1$): Vehicles flow freely on all streets. Notice that all platoons have a green light ahead of them. (B) \emph{Quasi-free-flow} phase ($\rho = 0.2$): Very few vehicles stop, and those that do only stop briefly. (C) \emph{Underutilized intermittent} subphase ($\rho = 0.3$): Some vehicles stop, there is some free space left at intersections. \emph{Full capacity intermittent} subphase ($\rho = 0.5$): Some vehicles stop, intersections are used at full capacity ($J=0.25$). Notice that all intersections are being utilized.}
     \label{fig:phases_so1}
\end{figure}

\begin{figure}
     \centering
     \subfigure[]{
          \label{fig:phases_so2A}
          \includegraphics[width=.45\textwidth]{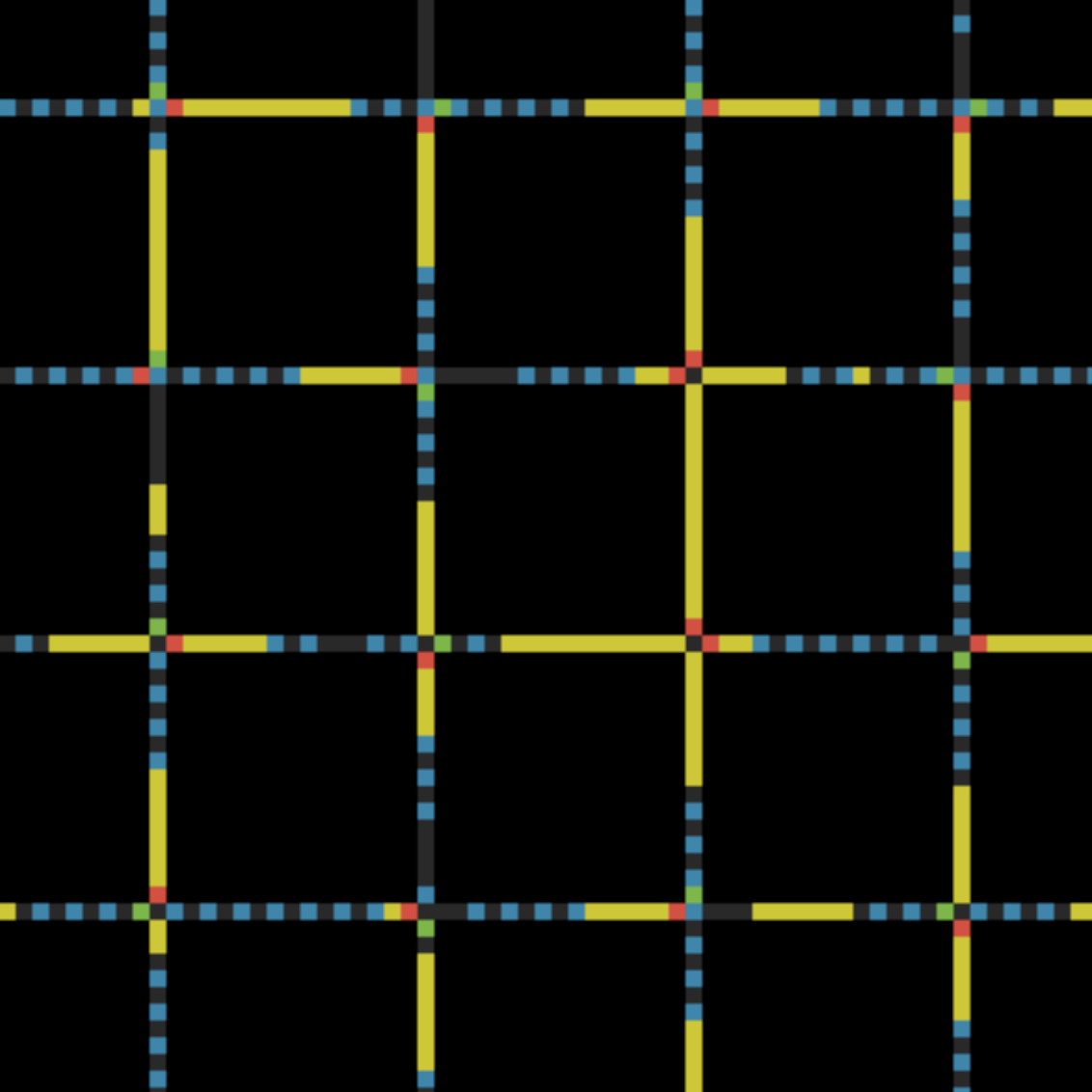}}
     \subfigure[]{
          \label{fig:phases_so2B}
          \includegraphics[width=.45\textwidth]{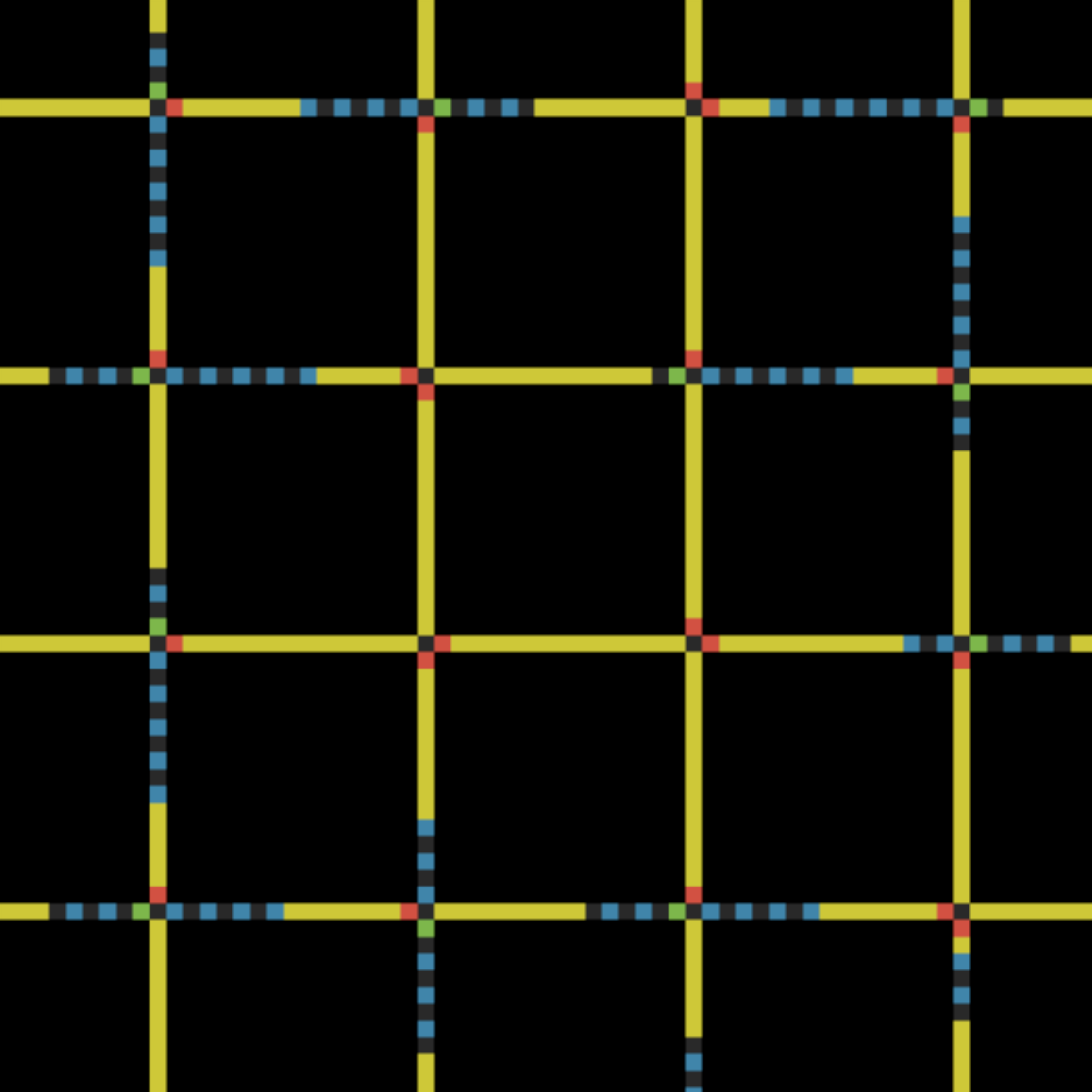}}
\\
     \subfigure[]{
          \label{fig:phases_so2C}
          \includegraphics[width=.45\textwidth]{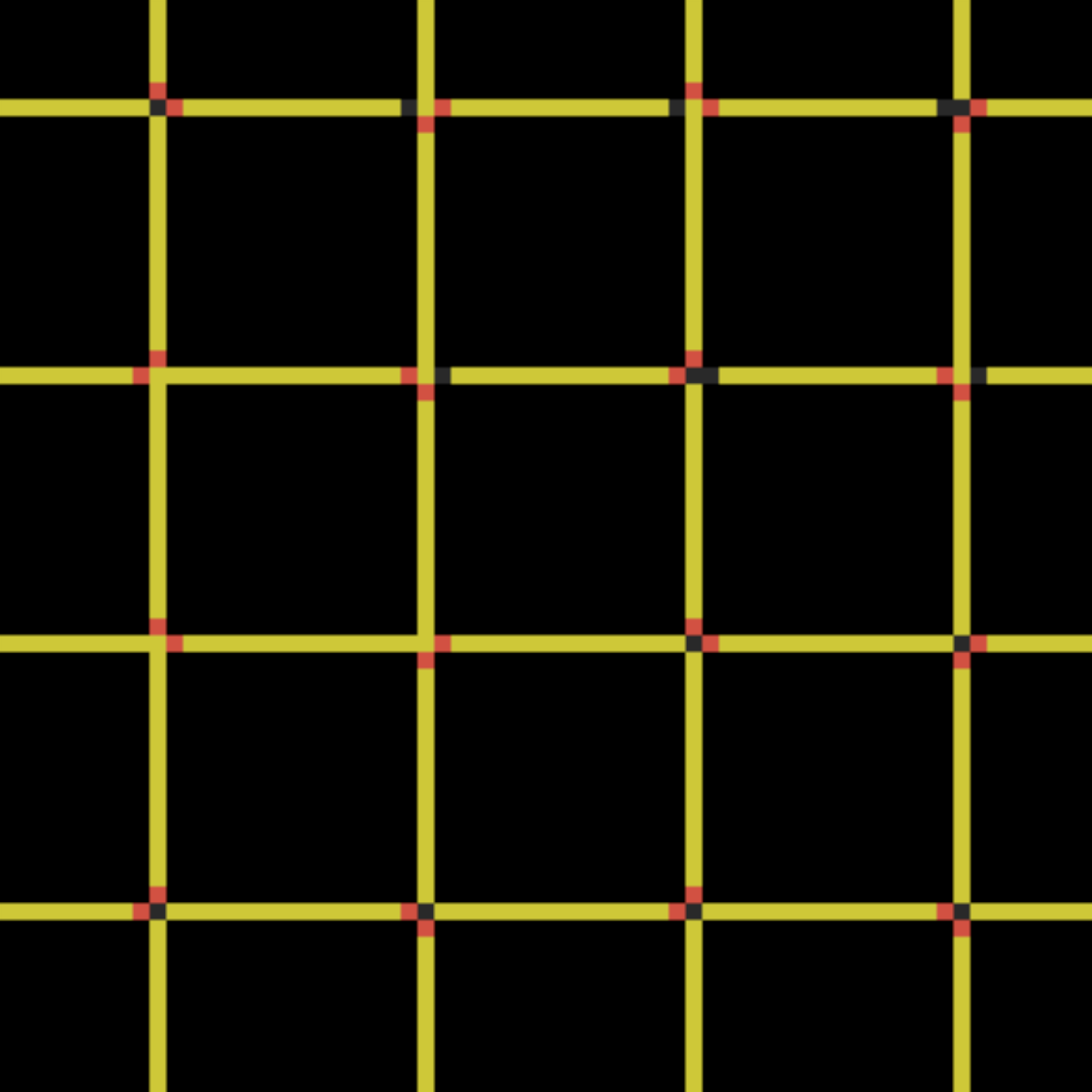}}
     \subfigure[]{
          \label{fig:phases_so2D}
          \includegraphics[width=.45\textwidth]{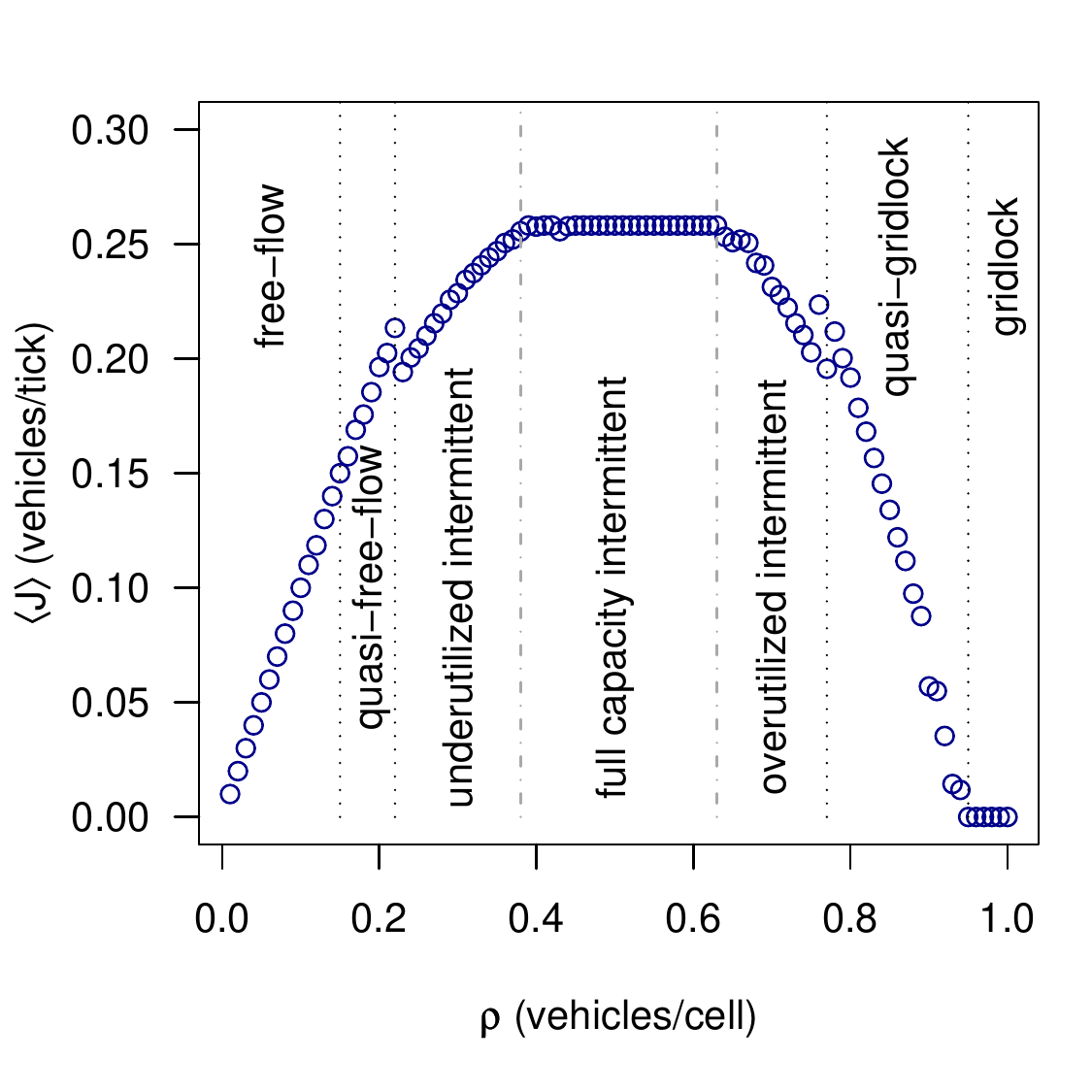}}

     \caption{Screenshots of different phases for \emph{self-organizing} method (A--C). Four-by-four sections of ten-by-ten simulations shown. Blue cells indicate moving vehicles, yellow cells indicate stopped vehicles. See also Figure \ref{fig:phases_so1}. (A) \emph{Overutilized intermittent} subphase ($\rho = 0.7$): Some vehicles stop, some intersections have to stop traffic in both directions due to high density. This subphase is analogous to the \emph{underutilized intermittent} subphase (Figure \ref{fig:phases_so1C}), where intersections are not used at their full capacity because of low density.
     (B) \emph{Quasi-gridlock} phase ($\rho = 0.8$): Almost all vehicles are stopped, but free spaces self-organize and flow in the opposite direction, triggering green lights when they are about to reach an intersection. Notice that all intersections with free space have a green light. This phase is analogous to the \emph{quasi-free-flow} phase (Figure \ref{fig:phases_so1B}), where vehicles self-organize in platoons and trigger green lights when they are about to reach an intersection.
     (C) \emph{Gridlock} phase ($\rho = 0.95$): At extremely high densities, initial conditions lead to a flow standstill ($v=0$). This phase is analogous to the \emph{free-flow} phase (Figure \ref{fig:phases_so1A}), where no vehicle has to stop ($v=1$).
     (D) Flux diagram (same as Figure \ref{fig:results_cityC}) indicating different phases.}
     \label{fig:phases_so2}
\end{figure}

There is a certain symmetry in the phases and in the flux diagram (Figures \ref{fig:results_cityC} and \ref{fig:results_cityD}) of the \emph{self-organizing} method. On one extreme (low density) there is the \emph{free-flow} phase, where no vehicle stops ($v=1$). On the other extreme (very high density) there is the \emph{gridlock} phase, where no vehicle moves ($v=0$). As the density moves towards the middle, on the low-density side there is formation of platoons that trigger green lights in the \emph{quasi-free-flow} phase. On the high-density side there is formation of free spaces that also trigger green lights in the \emph{quasi-gridlock phase}. For medium densities we have the \emph{intermittent} phase with its three subphases. In the center there is the \emph{full capacity} subphase, where there is always a vehicle crossing an intersection, leading to a maximum flux $J=0.25$. On the low-density side, the  \emph{underutilized} subphase cannot reach the full capacity because some intersections are idle. On the high-density side, the \emph{overutilized} subphase cannot reach full capacity because some intersections are stopping traffic from both directions. Screenshots of the different phases for the \emph{self-organizing} method can be seen in Figures \ref{fig:phases_so1} and \ref{fig:phases_so2}.

The average velocity over all densities in the experiments shown in Figure \ref{fig:results_cityB} for the \emph{green-wave} method was $\langle v\rangle \approx 0.22$. For the \emph{self-organizing} method it was $\langle v\rangle \approx 0.55$, i.e.\ a 150\% improvement over the \emph{green-wave} method. However, it is unrealistic to average over all densities, especially when the \emph{green-wave} method reaches a gridlock for 70\% of them. To have a better comparison, we can average the velocities for densities where both methods have not reached gridlocks, i.e.\ $\rho < 0.3$. The results for the \emph{green-wave} method were $\langle v\rangle_{\rho \leq 0.26} \approx 0.7$ and for the \emph{self-organizing} method $\langle v\rangle_{\rho \leq 0.26} \approx 0.95$, i.e.\ an improvement of 35\% over the \emph{green-wave} method, which is considerable. This is comparable with our previous results \citep{Gershenson2005,CoolsEtAl2007}.

As for the flux (shown in Figure \ref{fig:results_cityD}), the total average for the \emph{green-wave} method was $\langle J\rangle \approx 0.03$. For the \emph{self-organizing} method it was $\langle J\rangle \approx 0.18$. This is an unrealistic 470\% improvement over all densities. For lower densities, i.e.\ $\rho < 0.3$, $\langle J\rangle_{\rho \leq 0.26} \approx 0.09$ for the \emph{green-wave} method and $\langle J\rangle_{\rho \leq 0.26} \approx 0.12$ for the \emph{self-organizing} method, i.e.\ a 33\% improvement. Notice also that the maximum flux for the \emph{green-wave} method $J_{max} \approx 0.19$ (for a particular density) and for the \emph{self-organizing} method $J_{max}=0.25$ (for a broad range of densities), i.e.\ a 31\% improvement.

The \emph{green-wave} method is good for vehicles going in the
direction of the green wave at low densities. However, it performs poorly overall
because of the slow flow of vehicles going on the opposite
direction. The lights are anti-correlated in such a way that long
queues are formed for densities $\rho > 0.3$, blocking intersections
upstream and leading to gridlocks. Moreover, it has been shown that
this method is remarkedly sensitive to the value of $T$
\citep{BrockfeldEtAl2001}. Our \emph{self-organizing} method achieves
good performance for all densities compared with the green-wave method. It responds to the current traffic demand, so vehicles have to wait only if there are vehicles crossing at that moment. Rule 1 promotes the formation of platoons, which leave free space for other platoons to cross without interference. Together with rule 4, this achieves free-flow in four directions---for random initial conditions---for low densities. The performance at medium densities is also good, reaching the maximum possible flux for a broad range of densities. Traffic lights respond to the demand of platoons, and these do not have to wait for long, reducing the probability of long queues, which would interfere with intersections upstream. Rules 3, 5 and 6 alleviate the interference caused by long queues. Rule 3 prevents platoons from growing too much if there is a demand on the intersecting street, whereas rules 5 and 6 prevent vehicles from blocking an intersection upstream. Experiments varying the parameters of the \emph{self-organizing} method showed that the method is robust to changes in these values.

\subsection{Different city sizes}

\begin{figure}
     \centering
     \subfigure{
          \label{fig:results_xyA}
          \includegraphics[width=.45\textwidth]{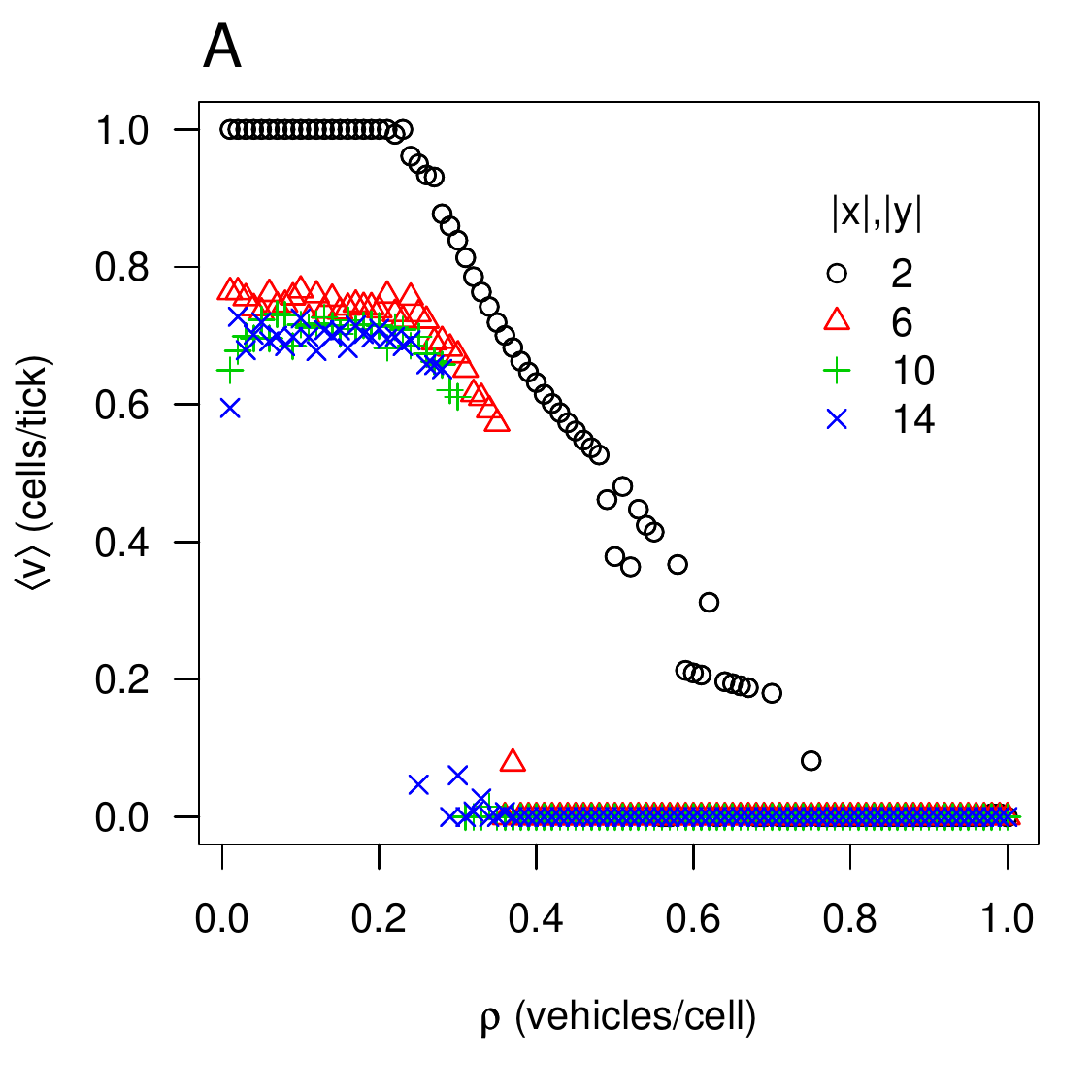}}
     \subfigure{
          \label{fig:results_xyB}
          \includegraphics[width=.45\textwidth]{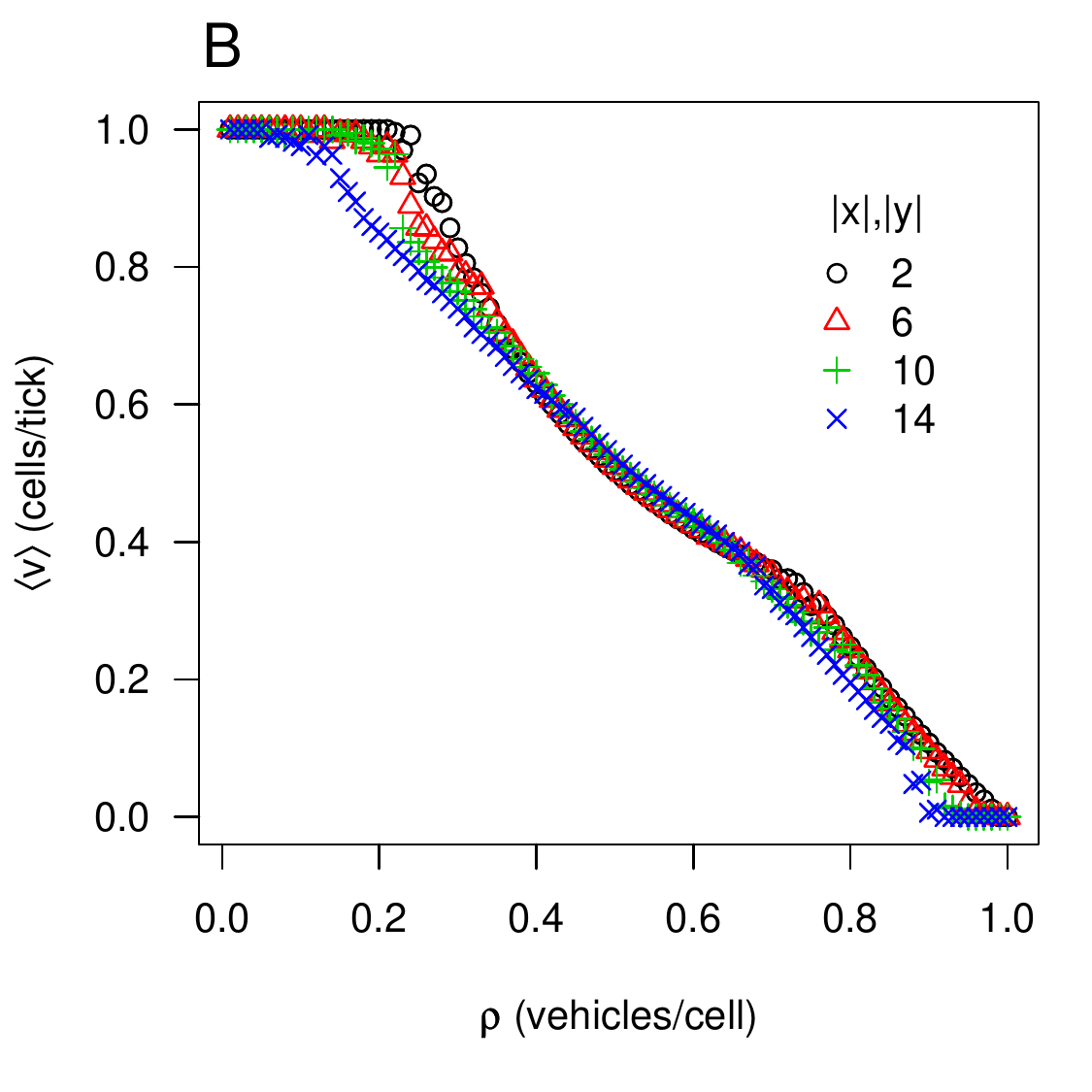}}
          \\
     \subfigure{
          \label{fig:results_xyC}
          \includegraphics[width=.45\textwidth]{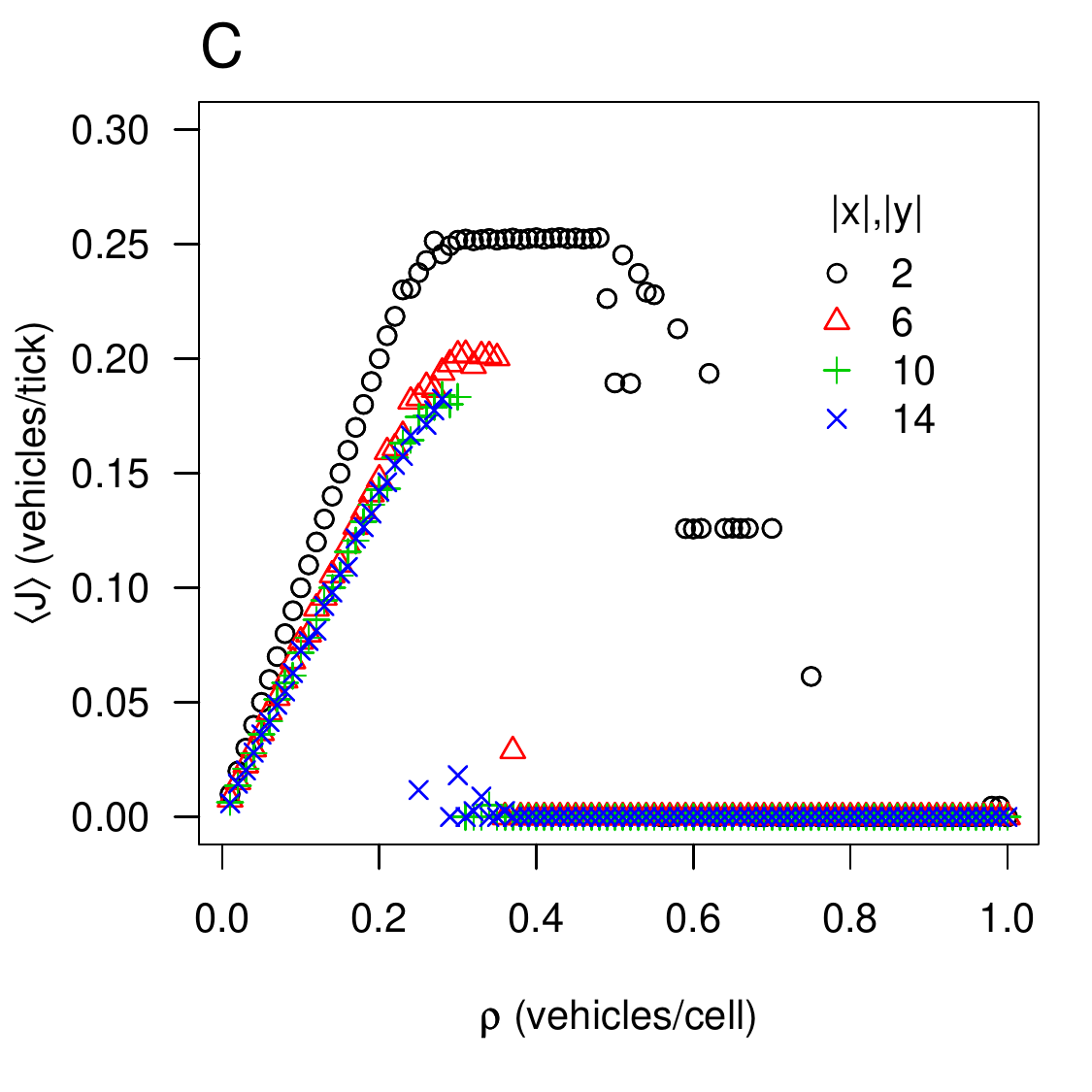}}
     \subfigure{
          \label{fig:results_xyD}
          \includegraphics[width=.45\textwidth]{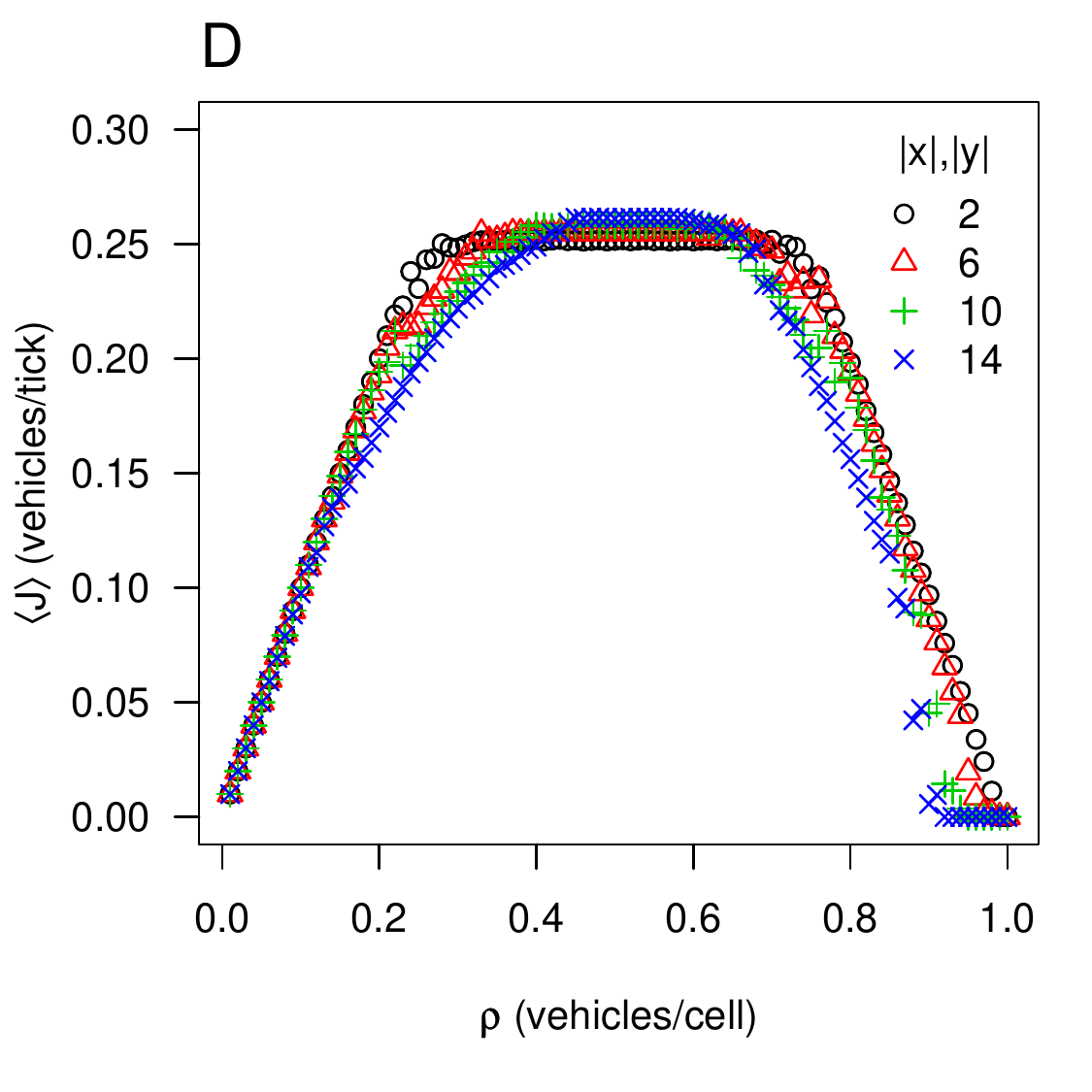}}

     \caption{Simulation results varying the number of streets in the $x$ and $y$ directions: (A,B) average velocity $\langle v\rangle$  and (C,D)  average flux $\langle J\rangle$ for different densities $\rho$: (A,C) \emph{Green-wave method} and (B,D) \emph{Self-organizing} method.}
     \label{fig:results_xy}
\end{figure}

To understand better the transition between the studied cases of one and one hundred intersections, we performed simulations varying the number of cyclic streets (all of 160 cells long). The results for two-by-two, six-by-six, ten-by-ten, and fourteen-by-fourteen configurations can be seen in Figure \ref{fig:results_xy}.

From the results we can see that both methods perform better for fewer intersections in the simulation. This is because it is easier to coordinate four intersections (two-by-two case, $|x|,|y|=2$) than 196 (fourteen-by-fourteen case, $|x|,|y|=14$). Actually, the \emph{green-wave} method can reach free-flow for the two-by-two case. Having only two intersections per street and cyclic boundaries makes the problem symmetric, and vehicles can flow freely in four directions (Figure \ref{fig:results_xyA}) and the system can reach a maximum flux $J=0.25$ (Figure \ref{fig:results_xyC}). However, even for the six-by-six case ($|x|,|y|=6$) the performance of the \emph{green-wave} method is quite poor, and decreases as the number of streets increases. For the \emph{self-organzing} method, the performance also decreases, although increasing the number of streets does not change the dynamical properties of the method. It is clear that when $|x|,|y|\rightarrow \infty$ there is no free-flow phase. 
Further simulations with very large grids ($|x|,|y|>100$) should be made to better understand this behavior. Will the performance decrease constantly? Or will it converge to a limit when $|x|,|y|\rightarrow \infty$? Other relevant experiments would be with open boundary conditions. In any case, real cities have a limited number of cities, but it is necessary to know how dependent the performance of traffic controllers is on the city grid size.
In Figure \ref{fig:results_xyD} the maximum capacity of the system seems to increase with $|x|,|y|$. This is an artifact of the model, since a vehicle can go into the intersection when a vehicle crossing it in the perpendicular street just left it, i.e.\ there can be two vehicles in contiguous cells on different streets, but not on the same street. 

\begin{figure}
     \centering
     \subfigure{
          \label{fig:results_dA}
          \includegraphics[width=.45\textwidth]{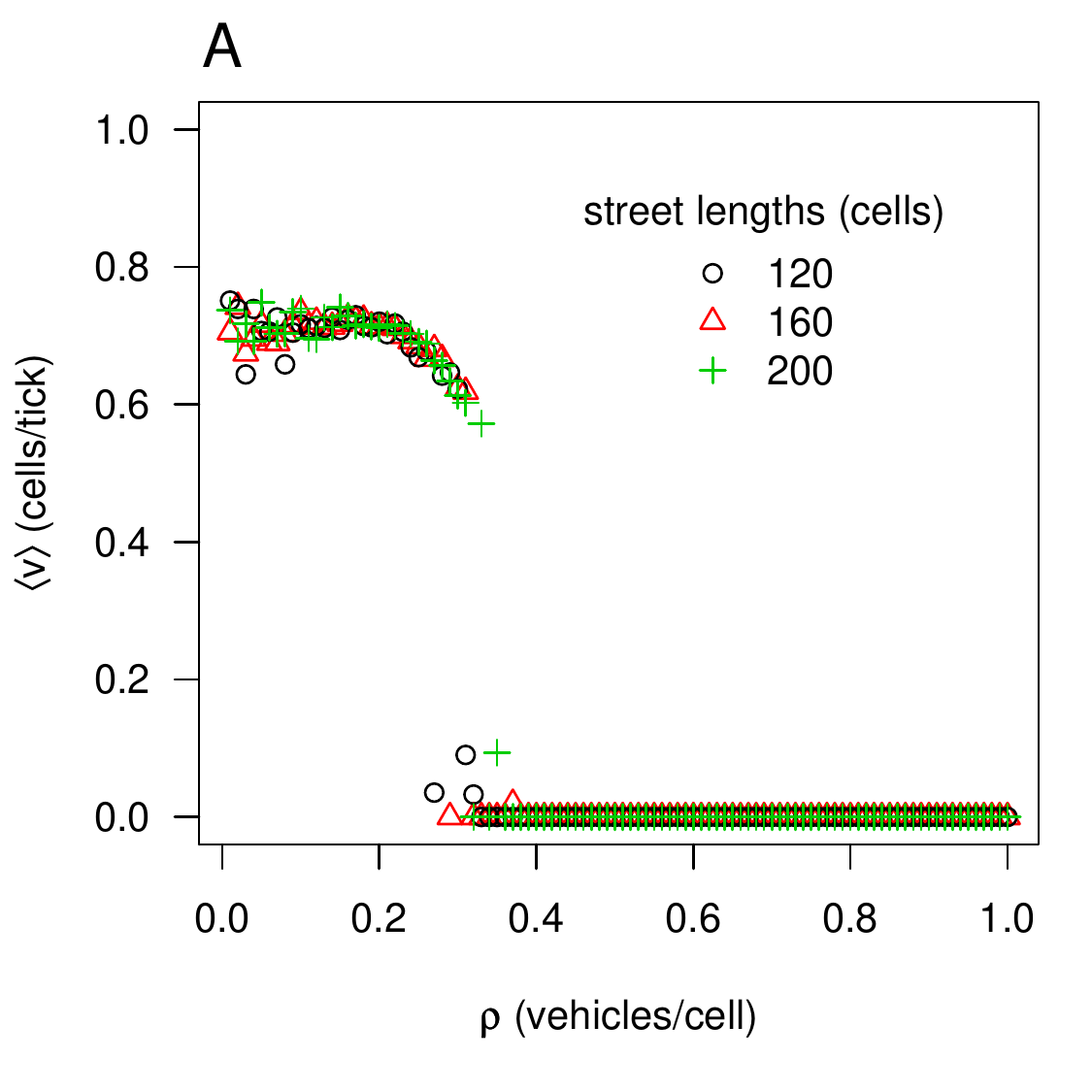}}
     \subfigure{
          \label{fig:results_dB}
          \includegraphics[width=.45\textwidth]{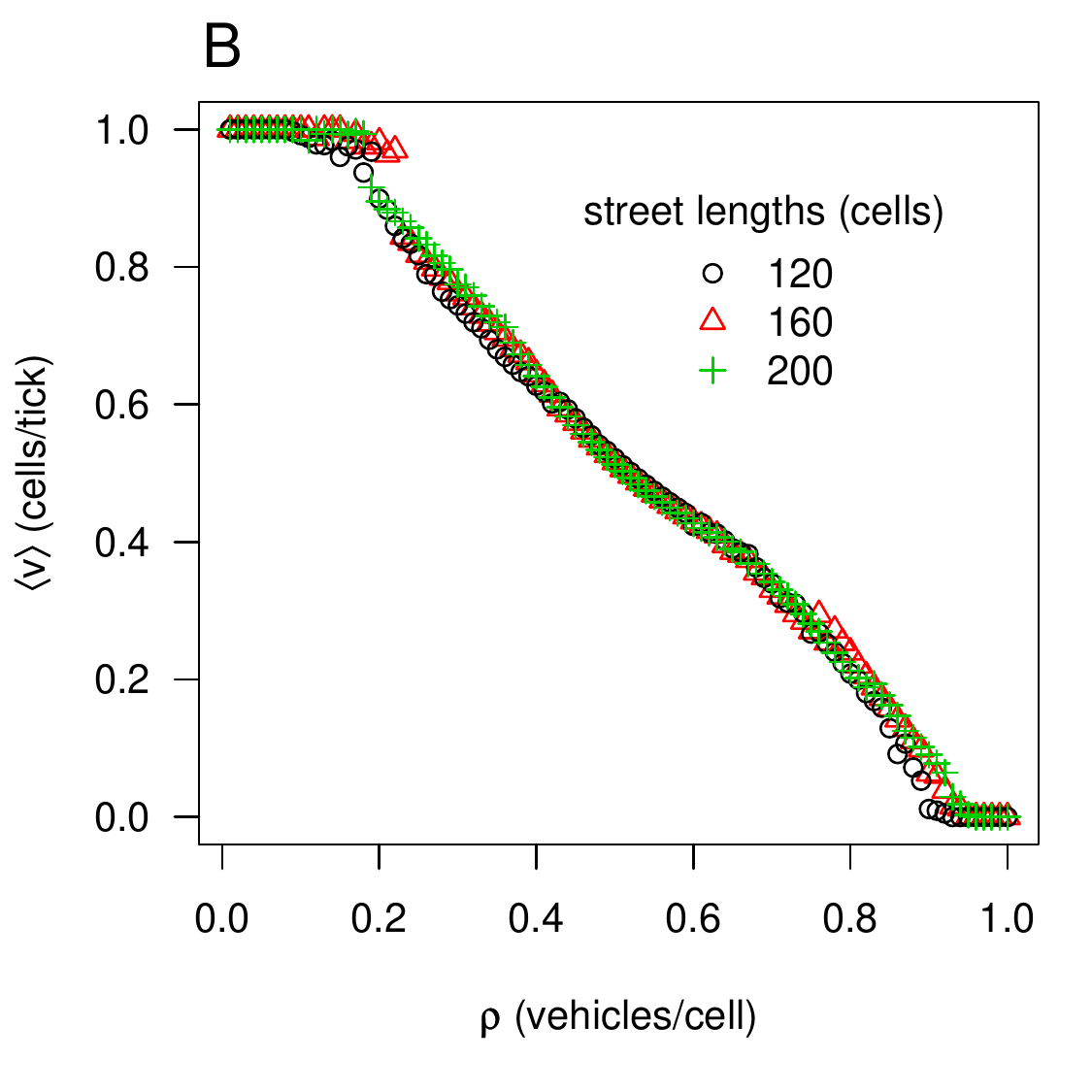}}
          \\
     \subfigure{
          \label{fig:results_dC}
          \includegraphics[width=.45\textwidth]{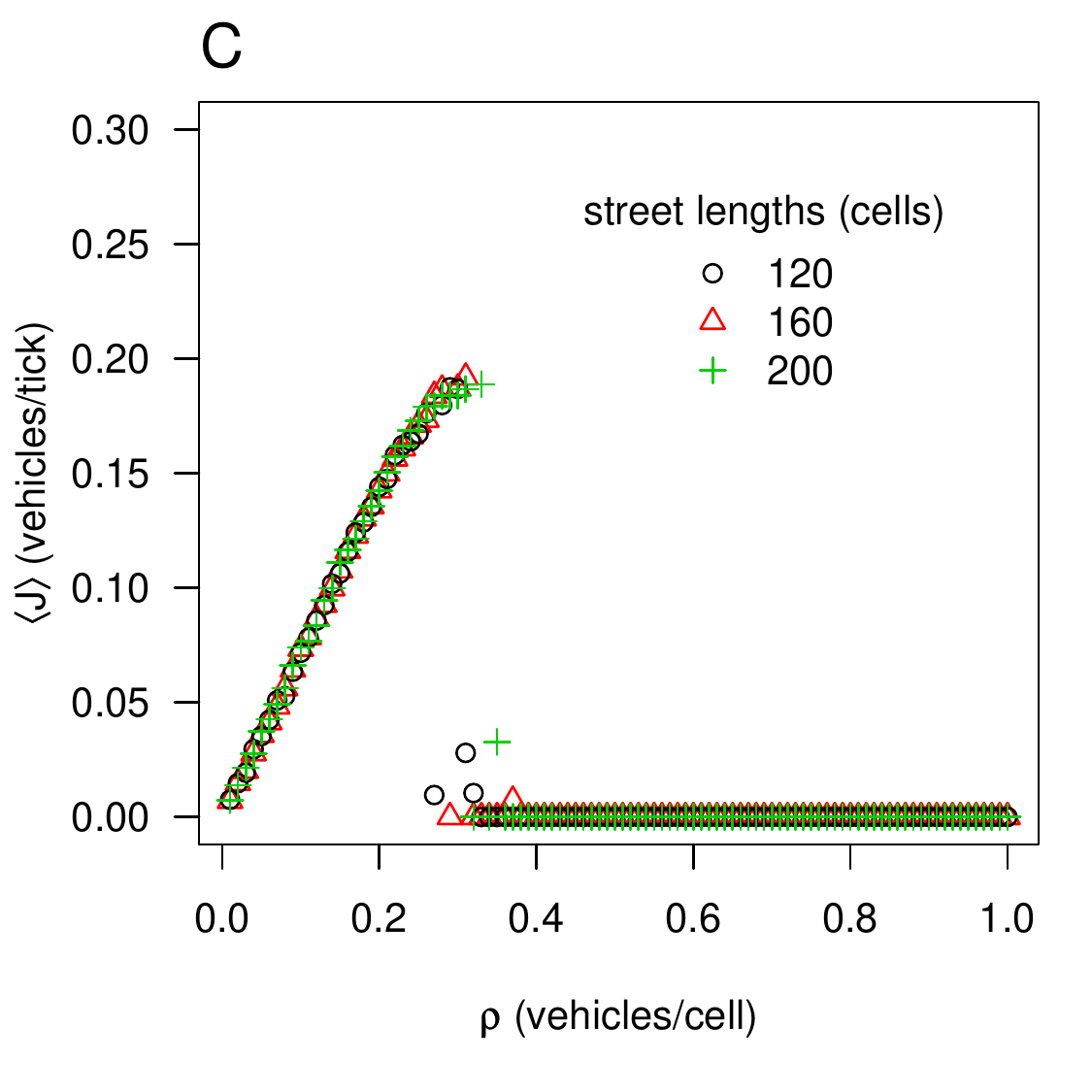}}
     \subfigure{
          \label{fig:results_dD}
          \includegraphics[width=.45\textwidth]{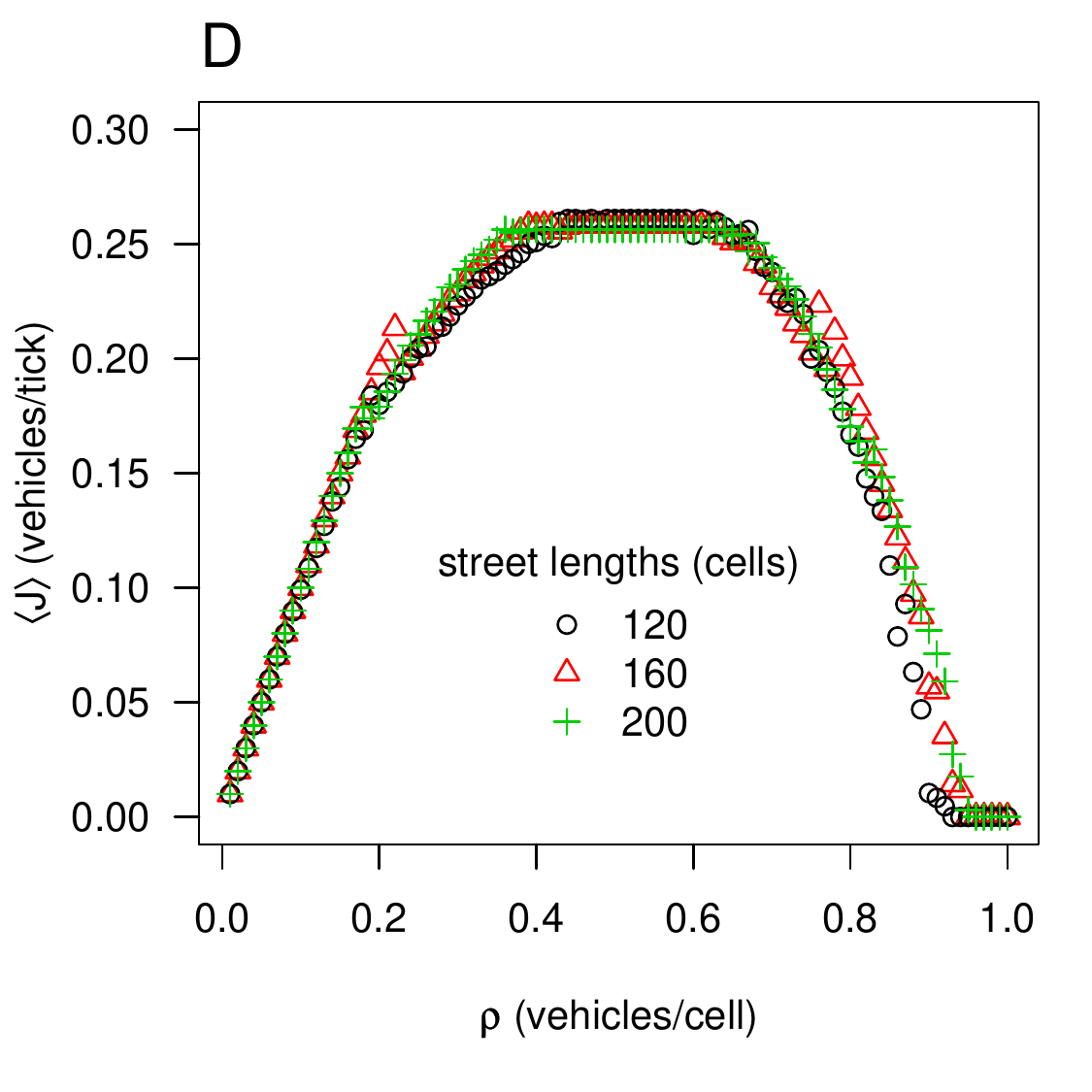}}

     \caption{Simulation results varying the street lengths in ten-by-ten scenarios: (A,B) average velocity $\langle v\rangle$  and (C,D)  average flux $\langle J\rangle$ for different densities $\rho$: (A,C) \emph{Green-wave method} and (B,D) \emph{Self-organizing} method.}
     \label{fig:results_d}
\end{figure}

Since in the above simulations all cities had a length of 160 cells, more intersections implied shorter interstreet distances. To check whether the variation on the performance was due to the number of intersections and not due to the interstreet distances, we performed simulations of three different ten-by-ten scenarios, varying the street lengths. Results are shown in Figure \ref{fig:results_d}. It can bee seen that the interstreet distances have no noticeable effect on the performance of both methods.

\section{Possible refinements}
\label{sec:refinements}

Our traffic model is as simple as possible. These are possible improvements that could be made to contemplate more realistic traffic situations:

\begin{itemize}
\item There are no amber lights in the current model. The behavior behind amber lights is equivalent to that behind red lights, i.e.\ vehicles should stop. This could be implemented in our model by adding a ``red-only'' phase, where both streets have a red light to allow clearing of the intersection. Actually, this is already implemented when rule 6 of the \emph{self-organizing} method demands both lights to turn to red.
This is achieved as follows: a) the cell previous to an intersection on a street with a light that is about to turn to red should change its rule ($184\rightarrow 252$) and b) the intersection should change its rule ($184\rightarrow 136$). During the red-only phase, the cell after the intersection remains using rule 184, to allow clearance of vehicles in the intersection in that direction. At the end of the red-only phase (and if the intersection is cleared): a) the cell after the intersection should change its value ($184\rightarrow 136$), b) the intersection and cells in the street that is turning green should change their rules to 184, and c) the intersection should change neighbors to those in the street with the green light.

\item Turns can be modeled with the rules already used in the model, but these should be rearranged to allow a turning vehicle to temporarily switch the rules between the cell right after the intersection in the street it was going before the turn ($184\rightarrow 136$) and the cell into which the vehicle will turn into ($136\rightarrow 184$). This change should be restored only when the turning vehicle leaves the intersection.

\item Multiple-lane streets could be modeled with parallel arrays of cellular automata, with further rules for lane changing, e.g.\ \citep{Knospe:2002}. This would also increase the number of cells that form intersections, so more changes should be made to ensure the clearance of vehicles on the direction they were heading.
\end{itemize}

As for the \emph{self-organizing} method to control traffic lights, this should be implemented in more realistic simulations and eventually in real cities. There are several aspects that our current results do not contemplate:

\begin{itemize}
\item The flux diagram of rule 184 (see Figure \ref{fig:results_streetC}) is symmetric, while it is well known that real highway traffic is not \citep{Hall:1986}. Similarly, we can expect that the quasi-symmetry of the flux diagram of the \emph{self-organizing} method will be affected in more elaborate simulations and in real traffic. However, the phases identified in this work can serve as useful guidance in further explorations.
\item The free-flow phase of the \emph{self-organizing} method is an artifact of the periodic boundary conditions and of the fact that no vehicle turns. In more realistic situations, there would be no free-flow phase. However, this would imply that there would be no gridlock phase (in theory) with open boundaries. Even if an intersection is blocked by e.g. an accident, the \emph{self-organizing} method would not allow the blockage to spread to other intersections by blocking flow into the affected intersection.
\item Pedestrians were not considered in our simulation. This is a necessary step before implementation, since pedestrians need about twenty seconds to cross safely a street, whereas the \emph{self-organizing} method can trigger faster switching times. Preliminary results that add five more rules to the \emph{self-organizing} method to include pedestrians have been encouraging and will be presented in future work.
\end{itemize}

\section{Conclusions}

There are several advantages of simple traffic models. Such models are easy to implement and reproduce and are computationally cheap. Also, by abstracting most details from real traffic, one can observe properties more clearly. For example, the phase transitions we found were not visible in our previous, more realistic multi-agent simulations. The phases can be identified, but the stochastic elements of those models smoothened the transitions, which are difficult to find analytically. With our simple model it is not possible to make realistic \emph{predictions}. However, it was possible to find better \emph{explanations} of why the \emph{green-wave} method is not efficient and why the upgraded \emph{self-organizing} method delivers such a great improvement. 

The optimal coordination of traffic lights is an EXP-complete problem \citep{PapadimitriouTsitsiklis1999}. Our results also showed that a successful alternative to optimization of complex problems lies in adaptation by self-organization. The \emph{self-organizing} method is able to coordinate traffic flows---not necessarily optimally, but efficiently---without the intersections having any prior knowledge of the incoming vehicles. This flexibility is a great advantage in such a complex problem domain.

The potential benefits of implementing the \emph{self-organizing} method are many. Improving traffic flow reduces the cost of transport, pollution and greenhouse gas emission, time lost during transit, and in general can improve the quality of life of citizens. However, caution should  be taken. On the one hand, an improved traffic flow could motivate more drivers to use a personal vehicle, perhaps even worsening the traffic conditions in the long run.
The solution to the traffic problem in cities is not only better controllers.  The main problem is that there are too many vehicles. Promoting alternative modes of transport and increasing the cost of personal vehicle usage are solutions that have been explored with different degrees of success in different scenarios. Nevertheless, our \emph{self-organizing} method can alleviate traffic problems while better alternatives are found.

\section*{Acknowledgments}

We gratefully acknowledge the facilities provided by IIMAS, UNAM.
Some of the improvements of the \emph{self-organizing} method over the one published previously \citep{Gershenson2005,CoolsEtAl2007} (rules 4 and 5) were developed by C. G. during a postdoctoral fellowship at the New England Complex Systems Institute, in collaboration with Yaneer Bar-Yam and Justin Werfel.


\bibliographystyle{cgg}
\bibliography{carlos,traffic,sos,complex,RBN}

\begin{appendix}

\section*{Appendix}

\section{\emph{Self-organizing} algorithm}
\label{App:Algo}

Algorithm \ref{alg:SOLA} formally describes our \emph{self-organizing} method.
Each intersection uses this algorithm independently to regulate traffic, i.e.\ there is no direct communication between intersections. The values of the parameters used in the simulations are shown in Table \ref{table:parameters}. We performed several simulations varying these parameters, and they are quite robust, i.e.\ the performance of the system is not affected by small changes in the parameters.

\incmargin{0.5cm}
\restylealgo{boxed}
\linesnumbered
\begin{algorithm}[h!]
\begin{footnotesize}
\Indp

\ForEach{$(\Delta t)$}{
	$t_{i}+= \Delta t$ \tcp*{local phase}

$k_{i}$ += $vehicles_{approachingRed}$ in $d$\tcp*{for rules 1 and 4}

\If {($vehicles_{stoppedAfterGreen}$ at $e > 0$) 
}{
      \If {($vehicles_{stoppedAfterRed}$ at $e > 0$) 	}{
      	$switchBothRed_{i}()$	\tcp*{rule 6}
      }
      \Else{      
      	$switchlight_{i}()$	\tcp*{rule 5}
	}
}

\ElseIf{$vehicles_{stoppedAfterRed}$ at $e == 0$}{
	
	  \If{bothRed?}{
		$restoreSingleGreen_{i}()$	\tcp*{complement to rule 6}
	  }	

	\If {($k_{i} \geq 1$) \textbf{and} ($vehicles_{approachingGreen}$ in $d == 0$)  }{
	      $switchlight_{i}()$	\tcp*{rule 4}	

	}

	\ElseIf {\textbf{not} ($0 < vehicles_{approachingGreen}$ in $r < m $)}{
		\tcp*{rule 3}
	 \If {($t _{i}\geq t_{\min}$) }{
		\tcp*{rule 2}
	    \If {($k_{i}$ $\geq $ $n $)}{
	      $switchlight_{i}()$ 	\tcp*{rule 1}
	    }
	   }
	  }
	}  
}

$switchlight_{i}()$ \Begin{
	$k_{i} = 0$;
	
	$t_{i} = 0$;
	
	 $switchTrafficLight_{i}()$;
	 	
}

  \caption{\emph{Self-organizing} method.}
\label{alg:SOLA}
\end{footnotesize}
\end{algorithm}
\decmargin{0.5cm}

On every tick ($\Delta t$), Algorithm \ref{alg:SOLA} increases the phase $t_{i}$  by the duration of $\Delta t$ (line 2), and the counter $k_{i}$ is increased by the number of vehicles approaching or waiting behind a red light within a certain distance $d$ (line 3). \emph{Rule 6} switches both lights to red if both streets are blocked ahead of the intersection (line 6). \emph{Rule 5} changes a green light to red if there are vehicles stopped ahead of green light at a distance $e$ from the intersection (line 9). This prevents the accumulation of vehicles when they cannot advance, diminishing the probability of their blocking the intersection, while at the same time allowing vehicles in the crossing street (if any) to advance. Rules 5 and 6 are normally used at high vehicle densities. A single green light is restored if there are no vehicles stopped ahead of the intersection and both lights are red (line 14). All of the following rules, to determine whether the light will switch, also depend on the condition that there is free space ahead of the red light.
With \emph{rule 4}, if there are no vehicles approaching a green light within a distance $d$, and there is at least one vehicle approaching a red light ($k_{i} \geq 1$), this is switched, so that by the time the vehicle(s) reach the intersection it(they) will not need to stop (line 17). This rule is normally used for low vehicle densities. 
\emph{Rule 3} prevents the ``tails" of platoons from being cut, by delaying the switching of a green light when there are few vehicles (fewer than $m$) just about to cross, i.e.\ within a distance $r$ (line 19). Still, rule 3 allows the division of long platoons, preventing the accumulation of vehicles waiting behind a red light. \emph{Rule 2} prevents the fast switching of traffic lights caused by high vehicle densities with a minimum phase $t_{\min}$ (line 20). If rules 2 and 3 are satisfied, \emph{rule 1} changes a traffic light when the count $k_{i}$ reaches a certain threshold $n$ (lines 21--22). This makes single vehicles wait for some time, increasing the probability that more vehicles will join them and thus promoting the formation of platoons. Once platoons reach a certain size, they can request a green light before reaching the intersection, if all other conditions are met. Even if this does not occur, once the conditions are proper, the vehicles will get a green light. Thus, in principle vehicles have to wait very little time because of red lights. When a traffic light is switched (lines 28--32), the counter $k_{i}$ and the phase $t_{i}$ are reset (lines 29--30). The phase $t_{i}$ keeps the time since the last light switch. Afterwards, the traffic lights are changed (line 31).

\begin{table}[htdp]
\begin{center}
\begin{tabular}{cccc}
\hline
	\textbf{Variable}&	\textbf{Abstract Value}&	\textbf{Scaled Value}&	\textbf{Used by}
\tabularnewline \hline
$\Delta t$	&1 tick		&1/3 s		&		Algorithm
\tabularnewline
$n$	&40 vehicles$\cdot$tick		&13.33 vehicles$\cdot$s		&	Rule 1	
\tabularnewline
$d$	&10 cells		&50 m		&Rules 1 and 4		
\tabularnewline
$t_{\min}$	&10 ticks		&3.33 s		&Rule 2		
\tabularnewline
$m $		&2 vehicles	&2 vehicles		&Rule 3		
\tabularnewline
$r$	&5 cells		&25 m		&Rule 3		
\tabularnewline
$e$	&2 cells		&10 m		&Rules 5	and 6	
\tabularnewline
\hline
\end{tabular}

\caption{Parameters used by \emph{self-organizing} method in simulations.}
\label{table:parameters}
\end{center}
\end{table}

\end{appendix}

\end{document}